\documentclass{aa} 

\usepackage{graphicx} 
\usepackage{epsfig}

\renewcommand{\d}{{\rm d}} 
\newcommand{\inta}{\int_{-i\infty}^{+i\infty}} 
\newcommand{\beq}{\begin{equation}} 
\newcommand{\eeq}{\end{equation}} 
\newcommand{\beqa}{\begin{eqnarray}} 
\newcommand{\eeqa}{\end{eqnarray}} 
\newcommand{\bea}{\begin{array}} 
\newcommand{\ea}{\end{array}} 
\newcommand{\cG}{{\cal G}} 
\newcommand{\rhob}{\overline{\rho}} 
 
\newcommand{\lag}{\langle} 
\newcommand{\rag}{\rangle} 
\newcommand{\Om}{\Omega_{\rm m}} 
\newcommand{\Ol}{\Omega_{\Lambda}} 
\newcommand{\Ob}{\Omega_{\rm b}} 
\newcommand{\xib}{\overline{\xi}} 
\newcommand{\tcool}{t_{\rm cool}} 
\newcommand{\tcoolef}{t_{\rm c,ef}} 
\newcommand{\theat}{t_{\rm heat}} 
 
\newcommand{\dR}{\delta_R} 
\newcommand{\rhop}{\rho_+} 
\newcommand{\bx}{{\bf x}} 
\newcommand{\rhoR}{\rho_R} 
\newcommand{\rhopm}{\rho_{\pm}} 
\newcommand{\rhom}{\rho_-} 
\newcommand{\rhoploc}{\rho_{+,\rm loc}} 
\newcommand{\rhopext}{\rho_{+,\rm ext}} 
\newcommand{\Tloc}{T_{\rm loc}} 
\newcommand{\Text}{T_{\rm ext}} 
\newcommand{\rhoc}{\rho_c} 
\newcommand{\dc}{\delta_c} 
\newcommand{\Tnl}{T_{\rm nl}} 
\newcommand{\rhomloc}{\rho_{-,\rm loc}} 
\newcommand{\rhomext}{\rho_{-,\rm ext}} 
\newcommand{\Ta}{T_{\alpha}} 
\newcommand{\nHI}{n_{\rm HI}} 
\newcommand{\nHII}{n_{\rm HII}}

\newcommand{\Tc}{T_c} 
\newcommand{\Tcol}{T_{\rm col}} 
\newcommand{\Tgh}{T_{\rm gh}}

\newcommand{\xiw}{\xi_{\rm w}} 
\newcommand{\br}{{\bf r}} 
\newcommand{\rhow}{\rho_{\rm w}} 
\newcommand{\Fm}{F_{\rm m}} 
 
\newcommand{\Fmw}{F_{\rm m,warm}} 
\newcommand{\Cw}{C_{\rm w}} 
\newcommand{\cP}{{\cal P}} 
\newcommand{\ypm}{y_{\pm}} 
\newcommand{\Dvir}{\Delta_{\rm vir}} 
\newcommand{\zform}{z_{\rm form}} 
\newcommand{\cN}{{\cal N}} 
 
\begin{document} 
 
  

\title{The phase-diagram of cosmological baryons}    
\author{P. Valageas\inst{1}, R. Schaeffer\inst{1}, and Joseph Silk\inst{2}}   
\institute{Service de Physique Th\'eorique, CEN Saclay, 91191 Gif-sur-Yvette, France 
\and Astrophysics, Denys Wilkinson Building, Keble Road, Oxford OX1 3RH, UK}  
\date{Received / Accepted } 
 
\abstract{ 
We investigate the behaviour of cosmological baryons at low redshifts $z \la 
5$ after reionization through analytic means. In particular, we study the 
density-temperature phase-diagram which describes the history of the 
gas. We show how the location of the matter in this $(\rho,T)$ diagram 
expresses the various constraints implied by usual hierarchical 
scenarios. This yields robust model-independent results which agree with 
numerical simulations. The IGM is seen to be formed via two phases: a 
``cool'' photo-ionized component and a ``warm'' component governed by 
shock-heating. We also briefly describe how the remainder of the matter is 
distributed over galaxies, groups and clusters. We recover the fraction of 
matter and the spatial clustering computed by numerical simulations. We also 
check that the soft X-ray background due to the ``warm'' IGM component is 
consistent with observations. We find in the present universe a baryon 
fraction of 7\% in hot gas, 24\% in the warm IGM, 38\% in the cool IGM, 9\% 
within star-like objects and, as a still un-observed component, 22\% 
of dark baryons associated with collapsed structures, 
with a relative uncertainty no larger than 30\% on these numbers.  
  } 
 
\maketitle

\section{Introduction} 
\label{Introduction}

As is well known, the mass of baryons embedded within stars or galactic disks 
 in the current universe is quite small since it yields a baryonic parameter 
 $\Omega_{\rm gal} \sim 0.002 - 0.006$ (e.g., \cite{Fuk1}) while standard 
 nucleosynthesis calculations give $\Ob \simeq 0.045$ (e.g., 
 \cite{Tyt1}). Therefore, most of the baryonic matter should lie in the 
 intergalactic medium. This agrees rather well with the fact that at higher 
 redshift one observes a large amount of mass in the Lyman-$\alpha$ forest 
 which consists of moderate density fluctuations ionized by the background UV 
 flux emitted by distant galaxies. Thus, one gets $\Omega_{\rm Ly \alpha} 
 \sim 0.01-0.05$ at $z\sim 3$ (e.g., \cite{Fuk1}). However, as noticed in 
 \cite{Cen1} the mass within the Lyman-$\alpha$ forest decreases with time 
 and at $z=0$ summing over all observed contributions one obtains $\Ob \la 
 0.011$ which falls short of the required total baryonic mass. Hence at $z=0$ 
 a large part of the baryons must lie in a new intergalactic component 
 beyond the usual Lyman-$\alpha$ forest clouds. As argued in \cite{Cen1} and 
 \cite{Dave2} this could be part of a ``warm'' phase of the intergalactic 
 medium (IGM), with temperatures in the range $10^5 < T <10^7$ K. 
 
The latter conclusion was reached from numerical simulations. In this 
article, we reconsider this problem in order to derive the properties of the 
IGM by analytic means. In particular, we wish to investigate whether one can 
understand this behaviour in a quantitative manner from robust, 
model-independent, arguments. 
 
First, in Sect.\ref{The phase-diagram of cosmological baryons} we study the 
$(\rho,T)$ phase-diagram of cosmological baryons. While the Lyman-$\alpha$ 
forest is described by a well-defined Equation of State the ``warm'' IGM 
component shows a broad scatter (e.g., \cite{Dave1}, \cite{Dave2}) since its 
temperature depends through shock-heating on the neighbouring gravitational 
potential which is a stochastic field. Nevertheless, we show that it is 
constrained to lie in a well-defined domain in the $(\rho,T)$ plane, and 
determine its average location in this plane, which may be considered as the 
``Equation of State'' of the ``warm'' IGM. We also give the location in this 
diagram of galaxies, groups and clusters. 
 
Next, in Sect.\ref{Distribution of matter} we use our results to compute the 
redshift evolution of the fraction of matter enclosed within the different 
phases. Then, in Sect.\ref{Warm IGM two-point correlation} we estimate the 
two-point correlation function and the clumping factor of the ``warm'' 
IGM. Finally, in Sect.\ref{Soft X-ray background} we check that the X-ray background emitted by the ``warm'' component agrees with observations.

\section{The phase-diagram of cosmological baryons} 
\label{The phase-diagram of cosmological baryons}

In this section we investigate the evolution of the physical properties of the baryons at moderate redshifts $z \la 5$, as gravitational clustering builds up. To do so, we study the distribution of matter in a density-temperature plane and determine the $(\rho,T)$ phase-diagram. Indeed, as we describe below baryons are located within specific regions in this $(\rho,T)$ plane which reflect the various processes (cooling, radiative heating, shock-heating, gravitational clustering,...) which take place in the universe. We display our results in Fig.\ref{diagrhoT}. Note that we use the overdensity $1+\delta=\rho/\rhob$ rather than the density $\rho$ in the figures, where $\rhob$ is the mean density of the universe.  
 
In this article, all our numerical results are obtained with the following cosmological parameters. We consider a low-density flat universe with $\Om=0.4$ and $\Ol=0.6$. The baryonic density parameter is $\Ob=0.0473$ and the Hubble constant is $H_0=65$ km/s/Mpc. The $\Lambda$CDM power-spectrum of the linear density fluctuations is normalized by $\sigma_8=0.8$.

\subsection{Temperature-scale relations} 
 
We consider two processes for the possible heating of the gas. First, the gas temperature may be governed by a local shock-heating due to gravitational processes. Thus, as large-scale structures build up we expect tidal effects and virialization processes to heat the gas within and in the neighbourhood of collapsed halos and filaments up to the virial temperature of the underlying gravitational potential well. The latter scales as $\Phi \sim \cG \rho R^2$, which defines the characteristic scale $R$, hence we write: 
\beq 
\mbox{local heating}: \;\;\; k T = \frac{8 \pi}{9} \mu m_p \cG \rho R^2 , 
\label{Tlocp1} 
\eeq 
where $\mu m_p$ is the mean molecular weight of the gas and $m_p$ the proton mass. 
 
Second, the gas temperature may be set by an external heating source, like a UV background radiation. Then, the temperature is no longer given by eq.(\ref{Tlocp1}) since it is not fixed by gravitational energy. However, the pressure of the gas erases the baryonic density fluctuations over a scale $R$ given by: 
\beq 
\mbox{external heating}: \;\;\; R = t_H C_s = t_H \sqrt{\frac{\gamma k T}{\mu m_p}} , 
\label{Rd1} 
\eeq 
where $C_s$ is the sound velocity and $\gamma=5/3$. This is simply the distance over which sound waves can travel in a Hubble time and build pressure equilibrium. Therefore, the gas sees the dark matter density field smoothed over this scale $R$.

There is a characteristic density $\rhoc$  at which shock heating starts to play a role, that is where eq.(\ref{Tlocp1}) is to be used in place of eq.(\ref{Rd1}). It is expected to correspond to a density contrast of a few units at least. The detailed discussion of the relation between the two regimes around $\rho \sim \rhoc$ may be found in Sect.\ref{WIGM} below.

\subsection{Exclusion constraints} 
\label{Exclusion constraints}

First, we note that the distribution of matter in the $(\rho,T)$ plane is strongly constrained by the existence of some exclusion regions where no particles can be found. The advantage of such conditions is that they are very robust in the sense they do not depend on the detailed history of the baryons. They apply to any hierarchical scenario of structure formation and provide the first guideline for the properties of baryons. We discuss these various constraints in the sections below. For sake of clarity, we turn within the diagrams shown in Fig.\ref{diagrhoT} in counter-clockwise order.

\subsubsection{Cooling constraint} 
\label{Cooling constraint}

The dashed-line on the far-right with an inclined ``V-shape'' is the familiar cooling constraint (see also \cite{Rees1}, \cite{Silk1}). It expresses the fact that high-density ionized hot gas ($T > 10^4$ K) cools very rapidly. Thus, baryons cannot remain for long in regions located to the right of this dashed-line. More precisely, we define the ``effective'' cooling time $\tcoolef$ set by atomic physics as: 
\beq 
\left( \frac{\d e}{\d t} \right)_{\rm cooling + UV} = - \frac{e}{\tcool} + \frac{e}{\theat} \equiv - \frac{e}{\tcoolef} , 
\label{ecool1} 
\eeq 
where $e=3/2 n_b k T$ is the specific energy and $n_b$ is the baryon number density. The cooling time $\tcool$ describes collisional excitation, collisional ionization, recombination, molecular hydrogen cooling, bremsstrahlung and Compton cooling or heating (e.g., \cite{Ann1}). The heating time $\theat$ is the radiative heating time due to a UV background radiation of the form $J_{\nu} \propto \nu^{-1}$. As usual, we normalize the UV flux by its value $J_{21}$ at the HI ionization threshold ($912 \AA$) in units of $10^{-21}$ erg s$^{-1}$ Hz$^{-1}$ cm$^{-2}$ sr$^{-1}$. The values we choose for $J_{21}$ are given in Tab.1 as a function of redshift. They are consistent with observations (e.g., \cite{Gial1}, \cite{Cook1}, \cite{Vog1}) and are the same as the ones used for modeling the evolution of Lyman-$\alpha$ clouds \cite{VLy}. Then, the cooling curve shown in Fig.\ref{diagrhoT} is given by the condition: 
\beq 
{\rm cooling:}  \;\;\; \tcoolef = t_H . 
\label{cool1} 
\eeq 
Note that in eq.(\ref{cool1}) we neglected shock-heating due to gravitational processes (e.g., virialization) and possible adiabatic cooling due to the expansion of the fluid elements. However, these processes are governed by the time $t_H$ which measures the expansion rate of the universe and the time-scale over which new non-linear structures form. Therefore, they do not modify the form of eq.(\ref{cool1}): they may only change the ratio between both sides of eq.(\ref{cool1}) (which we set equal to one) by a factor of order unity. Moreover, we have $\tcoolef \simeq \tcool$ as radiative heating does play a strong role here. Hence this constraint (\ref{cool1}) is actually quite robust. Then, the regions located to the right of the cooling curve have a cooling time which is less than the Hubble time. As a consequence, no particle can remain in this part of the $(\rho,T)$ plane for a Hubble time. This process is actually at the origin of galaxy formation as this gas cools and falls to the center of the gravitational well to form a disk and stars (see Sect.\ref{virialized halos} and \cite{Vgal}). This means that the matter which entered this region at some time should accumulate on the lower branch of the cooling curve (at $T \sim 10^4$ K) under the form of high-density cool gas which evolves on a Hubble time (but some of this gas also turns into stars). Note that this accumulation of baryons along the lower branch is clearly seen in the numerical simulations displayed in Fig.11 in \cite{Dave1} and in Fig.9 in \cite{Spr1}.

\begin{table} 
\begin{center} 
\caption{Redshift evolution of the background UV flux $J_{21}(z)$ used  
in this article.} 
\begin{tabular}{cccccc}\hline 
 
 $z=$0  &  1  &  2  &  3  &  4  &  5  \\ 
\hline \hline 
\\ 
$J_{21} =$ 0.05 & 0.5 & 0.8 & 0.4 & 0.2 & 0.1  
 
\end{tabular} 
\end{center} 
\label{table1} 
\end{table}

\subsubsection{High-density fluctuations} 
\label{High-density fluctuations}

A second constraint is set by the properties of the dark matter density field itself. Indeed, at a given scale $R$ beyond some threshold $\rhop(R)$ large densities have a negligible probability and only involve a very small amount of matter. In the linear regime  for instance, at large scales $R$, the probability distribution function (pdf) $\cP(\dR)$ of the density contrast $\dR$ over a spherical cell of radius $R$ is a Gaussian of variance $\sigma(R)$. Here we note as usual $\sigma(R)$ the rms linear density fluctuation and we assume Gaussian initial density fluctuations. In the non-linear regime, for highly non-Gaussian probabilities, the analogue of this upper threshold can also be calculated. It is given in App.\ref{Dark matter density field} where we detail our model for the pdf $\cP(\rhoR)$, where $\rhoR = 1+\dR$ is the overdensity. The high-density cutoff defined by eq.(\ref{rhpm1}) obeys the asymptotic behaviour (\ref{rhopR1}) in the quasi-linear and highly non-linear regimes. Note that this asymptotic behaviour  actually is model-independent. Indeed, in the limit $\xib \ll 1$ we recover the Gaussian cutoff implied by the initial conditions (where $\xib=\lag \dR^2 \rag$). On the other hand, in the limit $\xib \gg 1$ we recover the fact that most of the matter is enclosed within high-density halos of density contrast $\dR \sim \xib$ which occupy a small fraction of the volume ($\sim 1/\xib$) of the universe. Therefore, the constraint given by the high-density cutoff of the pdf $\cP(\rhoR)$ in the $(\rho,T)$ plane is quite robust and it applies to all hierarchical scenarios of structure formation.

We assume that the baryon density $\rho_b$ scales as the dark matter density through 
\beq 
\rho_b = \frac{\Ob}{\Om} \rho . 
\label{rhob1} 
\eeq 
This is valid even in the cooling region discussed above in Sect.\ref{Cooling constraint} if we consider also the cold baryons. We then can obtain an upper cutoff for the baryon overdensity $\rhop(R)$ at any given scale $R$. It will turn out to be convenient to express this cutoff $\rhop(R)$ in terms of a density-temperature relation, in order to draw its consequences for the phase-diagram $(\rho,T)$ of the IGM. To do so we simply need the relation $T(\rho,R)$.

Substituting eq.(\ref{Tlocp1}) into the relation $\rhop(R)$ derived in eq.(\ref{rhpm1}) we obtain a curve $\rhoploc(T)$. The subscript ``loc'' in the overdensity cutoff $\rhoploc$ refers to the fact that this is a ``local'' heating process. It is due to the gravitational interaction with neighbouring structures. This high-density cutoff $\rhoploc(T)$ corresponds to the dot-dashed curve shown in the diagrams in Fig.\ref{diagrhoT} which runs from $\delta \sim 5$ up to $\delta \sim 10^4$ and which crosses the cooling curve. 
 
Substituting eq.(\ref{Rd1}) into the relation $\rhop(R)$ obtained from eq.(\ref{rhpm1}) we get a curve $\rhopext(T)$. This corresponds in Fig.\ref{diagrhoT} to the short branch which runs upward from the curve $\rhoploc(T)$ at $\delta \sim 10$. Indeed, we note that for $\delta \ga 10$ the gas is located close to non-linear structures so that local shock-heating must be taken into account. However, for large densities we have $\Tloc > \Text$ since eq.(\ref{Rd1}) yields $\Text \sim \cG \rhob R^2$. In this case external heating plays no role. This is why we only plot the curve $\rhopext(T)$ up to the characteristic density contrast where $\Tloc = \Text$. At higher densities the threshold $\rhopext(T)$ becomes irrelevant. In fact, we see in Fig.\ref{diagrhoT} that the curve $\rhopext(T)$ due to external heating plays no role since at moderate densities $\rho \sim \rhob$ it is repelled to quasi-linear scales (see Sect.\ref{Transition to non-linearity}). 
 
Then, the region to the upper-right of the curves $\rhoploc(T)$ and $\rhopext(T)$ in the $(\rho,T)$ plane shown in Fig.\ref{diagrhoT} corresponds to rare high-density fluctuations which are located in the tail of the pdf $\cP(\rhoR)$. Therefore, there should be very few particles beyond these lines. Thus, this defines a second exclusion region. 
 
Note that at $z=0$ this constraint yields an upper bound $T_+ \sim 10^{7.3}$ K $\simeq 1.7$ keV. Of course, there exist some halos with a larger temperature: massive X-ray clusters. However, these are rare objects which only contain a small fraction of the baryonic matter content of the universe (typically $10\%$) and they indeed correspond to the high-mass tail of the mass function. This is obviously consistent with the description of the baryonic matter which is worked out in this article, see Sect.\ref{virialized halos}.

\subsubsection{Transition to non-linearity} 
\label{Transition to non-linearity}

A  characteristic scale which enters the problem we investigate here is set by the transition to the non-linear regime. At a given epoch, we define this scale $R_0(z)$ by the relation: 
\beq 
\sigma(R_0,z) \equiv 1 . 
\label{R01} 
\eeq 
Thus, scales larger than $R_0$ are still within the linear regime. Then, there can be no shock-heating due to gravitational clustering on these large scales which have not turned non-linear yet. Using eq.(\ref{Tlocp1}) we obtain a characteristic temperature $\Tnl(\rho)$: 
\beq 
\mbox{transition to non-linearity}: \;\;\; k \Tnl = \frac{8 \pi}{9} \mu m_p \cG \rho R_0^2 , 
\label{Tnl1} 
\eeq 
which describes the transition to non-linear scales (which lie at $T \ll \Tnl(\rho)$). This is shown by the straight dashed-line ($\Tnl \propto \rho$) plotted in Fig.\ref{diagrhoT}. Besides, we note that known sources of external heating (e.g., the UV background radiation) cannot heat the IGM up to such high temperatures. Therefore, no fluid element can be located in the region to the upper-left of the curve $\Tnl(\rho)$ since no physical process which is active on these large scales can heat the gas to these high temperatures. This yields a third exclusion region.

\subsubsection{Low-density fluctuations} 
\label{Low-density fluctuations}

Finally, a fourth constraint on the distribution of matter is given by the low-density cutoff of the pdf $\cP(\rhoR)$ of the dark matter density field. This is the analog of the high-density cutoff discussed in Sect.\ref{High-density fluctuations}. The low-density cutoff $\rhom(R)$ is derived in App.\ref{Dark matter density field}, from eq.(\ref{rhpm1}). It obeys the asymptotic behaviours (\ref{rhomR1}). In the quasi-linear regime we again recover the usual Gaussian cutoff $\delta_- \sim -\sigma$. In the highly non-linear regime the small overdensity $\rhom \sim \xib^{\;-\kappa/2} \ll 1$ expresses the formation of extreme underdensities on small scales. Contrary to the high-density cutoff it is somewhat model-dependent through the exponent $\kappa$ but this has no strong effect on the $(\rho,T)$ phase-diagram (we typically have $\kappa \sim 0.8$). As in Sect.\ref{High-density fluctuations} we need a relation $T(R)$ in order to derive a condition of the form $\rhom(T)$. We again consider both cases of local and external heating, described by eq.(\ref{Tlocp1}) and eq.(\ref{Rd1}). This yields the curves $\rhomloc(T)$ and $\rhomext(T)$ shown by the two steep parallel dashed-lines in Fig.\ref{diagrhoT}, at $\rhoR \sim 10^{-1}$. The curve associated with external heating is the left one (i.e. lower densities or higher temperature), as can be seen from eq.(\ref{Tlocp1}) and eq.(\ref{Rd1}). Thus, this defines a fourth exclusion region to the left of these curves.

\subsection{Equation of State of the Lyman-$\alpha$ forest} 
\label{LyEOS}

Thus, so far we have obtained constraints on the distribution of matter in the $(\rho,T)$ plane by drawing four exclusion regions. This already gives quite useful information about the properties of the IGM which are very robust. Now, we investigate a different point, seeking the location of the gas in the $(\rho,T)$ plane. This amounts to deriving  an Equation of State for this component. 
 
As shown in \cite{Gne1} the low-density photo-ionized IGM exhibits such an Equation of State as the gas follows a specific relation $\Ta(\rho)$ with a rather small scatter. This was derived in \cite{Gne1} from the Zel'dovich approximation (\cite{Zel1}) which applies up to the moderately non-linear regime ($\xib \la 1$). Here we reconsider this problem and we show that this Equation of State is rather robust with respect to the past history of the gas and applies independently of  the validity of the Zel'dovich approximation. First, we assume photo-ionization equilibrium (we restrict ourselves to $z \la 5$ after reionization) and we only take into account Hydrogen. Therefore, the ionization equilibrium reads: 
\beq 
\Gamma \; \nHI = \alpha(T) \; \nHII n_e , 
\label{phot1} 
\eeq 
where $\Gamma$ is the photo-ionization rate and $\alpha(T)$ is the recombination rate. They are given by: 
\beq 
\Gamma = \int 4\pi J_{\nu} \sigma_{\rm HI} \frac{\d\nu}{h\nu} = 3.08 \times 10^{-12} \; J_{21}(z) \; \mbox{s}^{-1} 
\eeq 
and: 
\beq 
\alpha(T) = \alpha_J \left( \frac{T}{T_J} \right)^{-(\nu-1)} , 
\eeq 
where we defined: 
\beq 
\nu=1.7 , \; \alpha_J =  1.23 \times 10^{-13} \; \mbox{cm}^3 \mbox{s}^{-1} , \; T_J = 5.8 \times 10^4 \mbox{K}. 
\label{TJ1} 
\eeq 
The temperature $T_J$ we introduced in eq.(\ref{TJ1}) is the characteristic temperature reached by the gas through the heating due to the UV background radiation flux. It is given by: 
\beq 
k T_J \equiv \frac{ \int 4\pi J_{\nu} \sigma_{\rm HI}   (h\nu-h\nu_{\rm HI}) \frac{\d\nu}{h\nu} } { \int 4\pi J_{\nu} \sigma_{\rm HI} \frac{\d\nu}{h\nu} } \; (\simeq 5 \; \mbox{eV}) , 
\label{TJ2} 
\eeq 
where $h\nu_{HI}=13.6$ eV is the Hydrogen ionization threshold. Note that this temperature $T_J$ does not depend on the amplitude $J_{21}$ of the  UV background. Moreover, it is fixed by atomic physics, independently of cosmological parameters. Next, the temperature $T$ of a given fluid element evolves as: 
\beq 
\frac{1}{T} \frac{\d T}{\d t} = \frac{2}{3} \frac{1}{\rho} \frac{\d \rho}{\d t} + \frac{1}{\theat} 
\label{TJ3} 
\eeq 
where $\d/\d t$ is the Lagrangian time derivative. The heating time $\theat$ is given by: 
\beqa 
\frac{3/2 n_b k T}{\theat} & = & \int 4\pi J_{\nu} \sigma_{\rm HI} \nHI (h\nu-h\nu_{\rm HI}) \frac{\d\nu}{h\nu} \nonumber \\ & = & k T_J \nHI \Gamma , 
\label{theat1} 
\eeqa 
where we take $n_b=2 \rho_b/m_p$ since we approximate the gas as fully ionized Hydrogen. The evolution eq.(\ref{TJ3}) is the same as eq.(\ref{ecool1}), except that we neglect cooling (which is justified here since we consider here moderate densities and temperatures) and gravitational shock-heating which is irrelevant. On the other hand, the term $\d\rho/\d t$ represents the pressure work, which takes into account the expansion of the fluid element. Substituting eq.(\ref{theat1}) into eq.(\ref{TJ3}) we obtain: 
\beq 
\frac{1}{T} \frac{\d T}{\d t} = \frac{2}{3} \frac{1}{\rho} \frac{\d \rho}{\d t} + \frac{1}{t_{10}} \frac{\rho}{\rho_J} \left( \frac{T}{T_J} \right)^{-\nu} , 
\label{TJ4} 
\eeq 
where we define: 
\beq 
\rho_J = 3 \frac{\Om}{\Ob} \frac{m_p}{\alpha_J t_{10}} \simeq 1.1\times 10^{-28} \; \mbox{g cm}^{-3}  
\label{rhoJ1} 
\eeq 
and: 
\beq 
t_{10} = 10^{10} \; \mbox{years} . 
\eeq 
In eq.(\ref{TJ4}) we used the fact that Hydrogen is almost entirely ionized at the low redshifts which we consider here ($z < 5$, after reionization), as shown by the Gunn-Peterson test. Next, it is convenient to introduce the adiabat $K$ defined as the exponential of the specific entropy $s$. More precisely, we define $K$ and $s$ by: 
\beq 
K \equiv \frac{T}{T_J} \left( \frac{\rho}{\rho_J} \right)^{-2/3} , \;\; s \equiv \ln K . 
\label{s1} 
\eeq 
Then, eq.(\ref{TJ4}) may be written: 
\beq 
K^{\nu-1} \frac{\d K}{\d t} = \frac{1}{t_{10}} \left( \frac{\rho}{\rho_J} \right)^{1-2\nu/3} . 
\label{s2} 
\eeq 
Here we note that $1-2\nu/3 \simeq -0.13$ is a small number. Therefore, the specific entropy of the fluid shows a weak dependence on the evolution of its density $\rho$. Hence we approximate the solution of eq.(\ref{s2}) by: 
\beq 
K^{\nu} \simeq \nu \left( \frac{\rho}{\rho_J} \right)^{1-2\nu/3} \; \frac{t}{t_{10}} , 
\label{s3} 
\eeq 
where $\rho$ is the density at the time $t$ we consider. In terms of the temperature-density relation, this yields: 
\beq 
\Ta(\rho) =  T_J \left( \nu \frac{t}{t_{10}} \frac{\rho}{\rho_J} \right)^{1/\nu} , 
\label{Ta1} 
\eeq 
where we used eq.(\ref{s1}). Thus, the expression (\ref{Ta1}) gives the first-order term for the expansion of $\ln T$ in terms of the small parameter $(1-2\nu/3) (\d \ln\rho/\d \ln t)$. From eq.(\ref{rhoJ1}) we note that at $z \la 5$ we have $\rho \sim \rho_J$ and $t_H \sim t_{10}$ hence the temperature of the photo-ionized IGM will be of order $T_J \sim 10^4$ K.  
 
Note that eq.(\ref{Ta1}) is independent of the normalization of the UV flux $J_{\nu}$. Indeed, the efficiency of radiative heating is proportional to $J_{\nu} \nHI$ but the density of neutral Hydrogen scales as $1/J_{\nu}$ (at ionization equilibrium for almost fully ionized gas) so that $J_{\nu}$ cancels out. Therefore, the result (\ref{Ta1}) is quite robust since it does not depend on the value of the UV flux. In particular, eq.(\ref{Ta1}) still holds even if the UV background is inhomogeneous: we only need to assume local ionization equilibrium. This explains why there is only a very small scatter around the Equation of State (\ref{Ta1}) since the actual physical conditions within these small clouds are almost independent of the actual history of each fluid element (i.e. the evolution of its density and local UV flux). This indeed agrees with the results of numerical simulations (e.g., \cite{Dave1}). 
 
\begin{figure*} 
\begin{center} 
\epsfxsize=8.1 cm \epsfysize=6 cm {\epsfbox{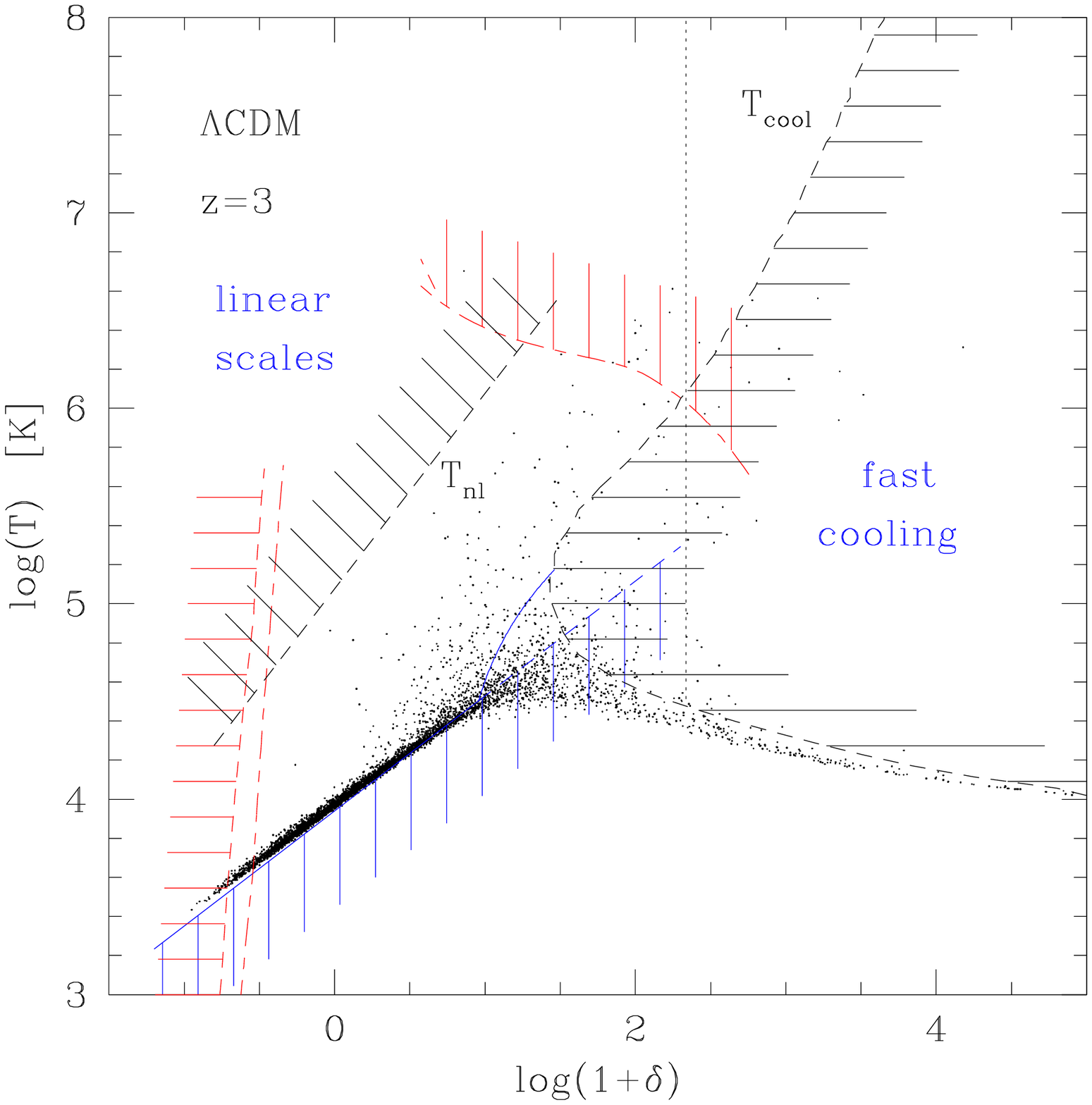}} 
\epsfxsize=8.1 cm \epsfysize=6 cm {\epsfbox{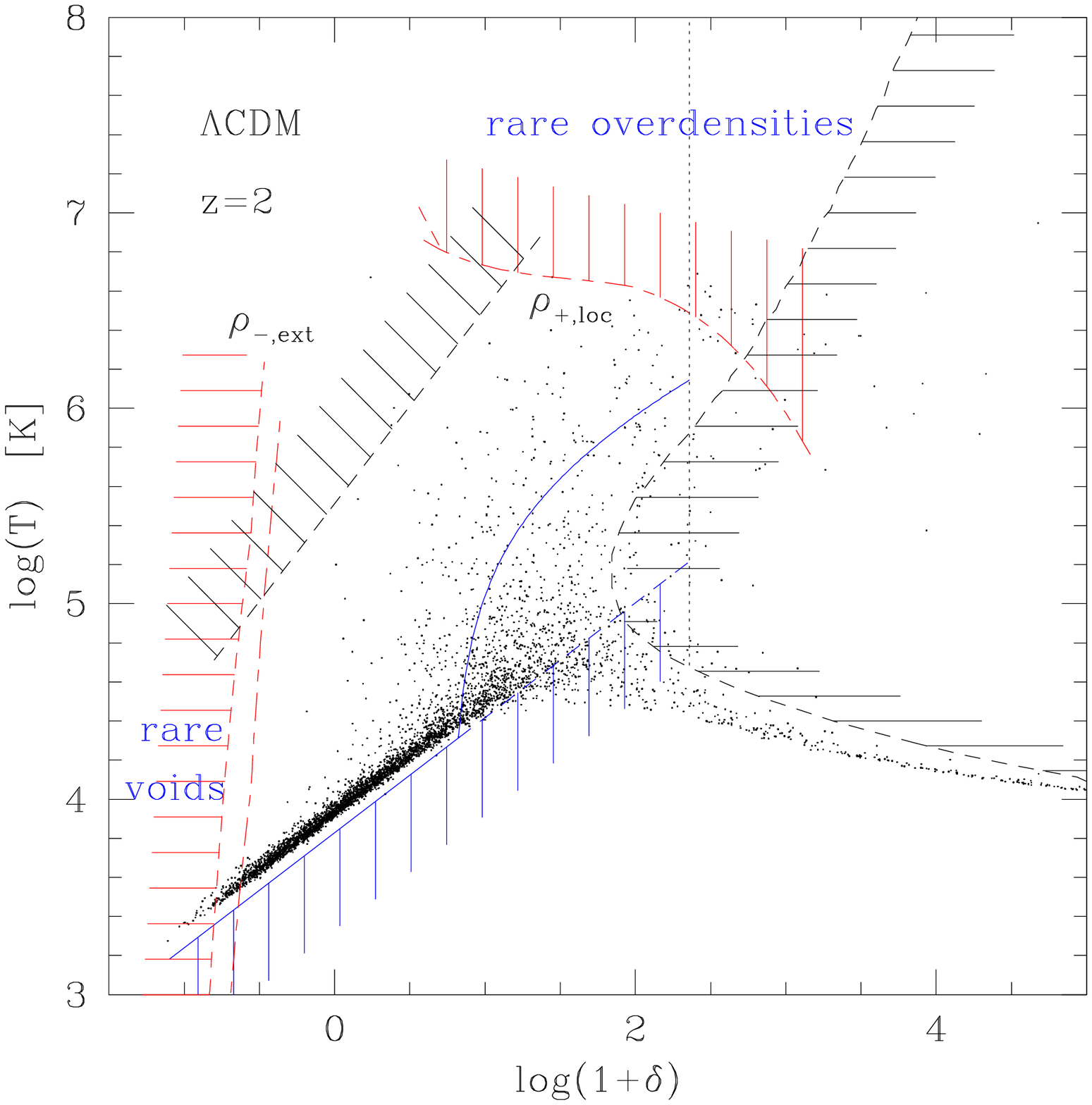}}\\ 
\epsfxsize=8.1 cm \epsfysize=6 cm {\epsfbox{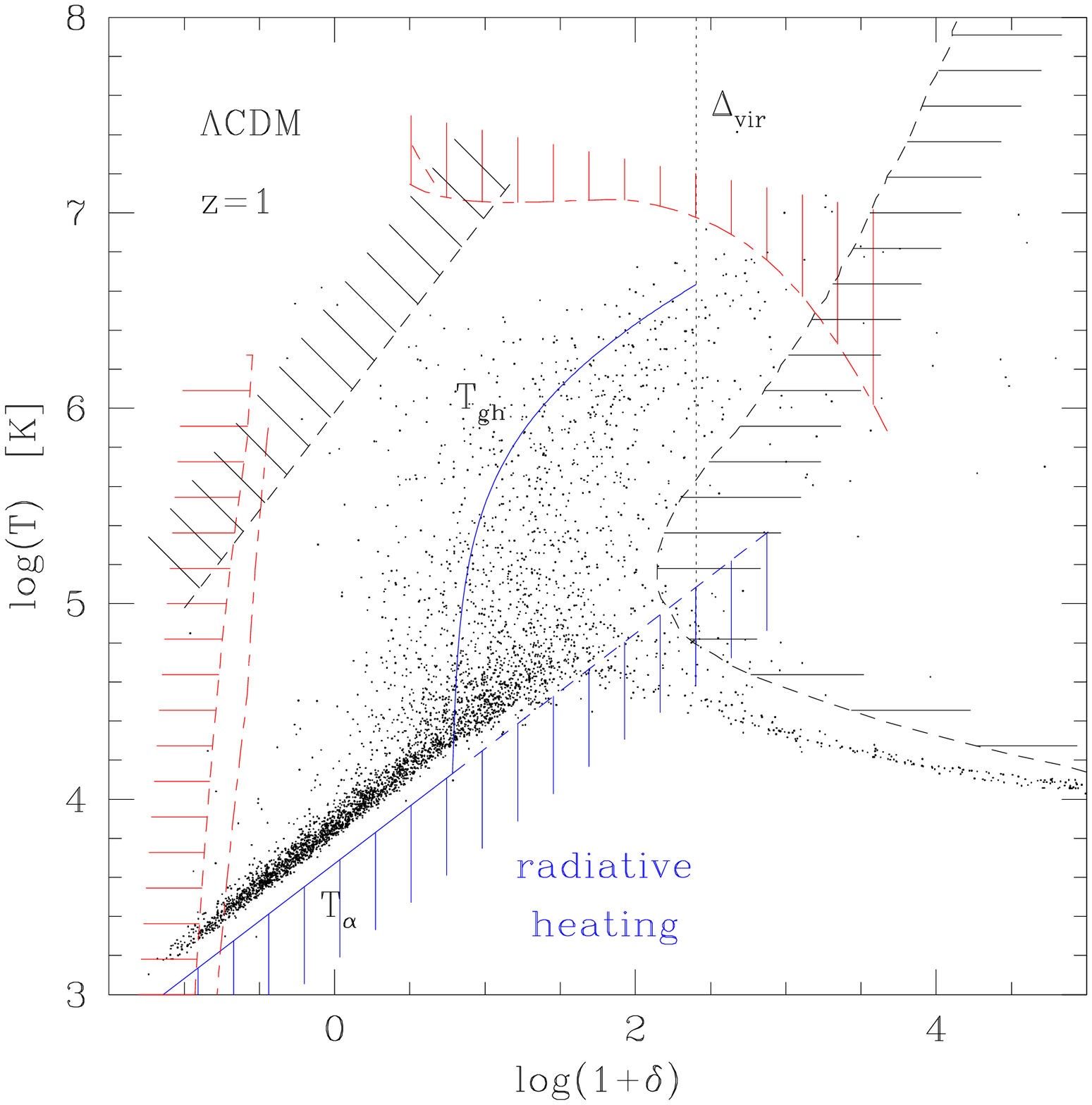}} 
\epsfxsize=8.1 cm \epsfysize=6 cm {\epsfbox{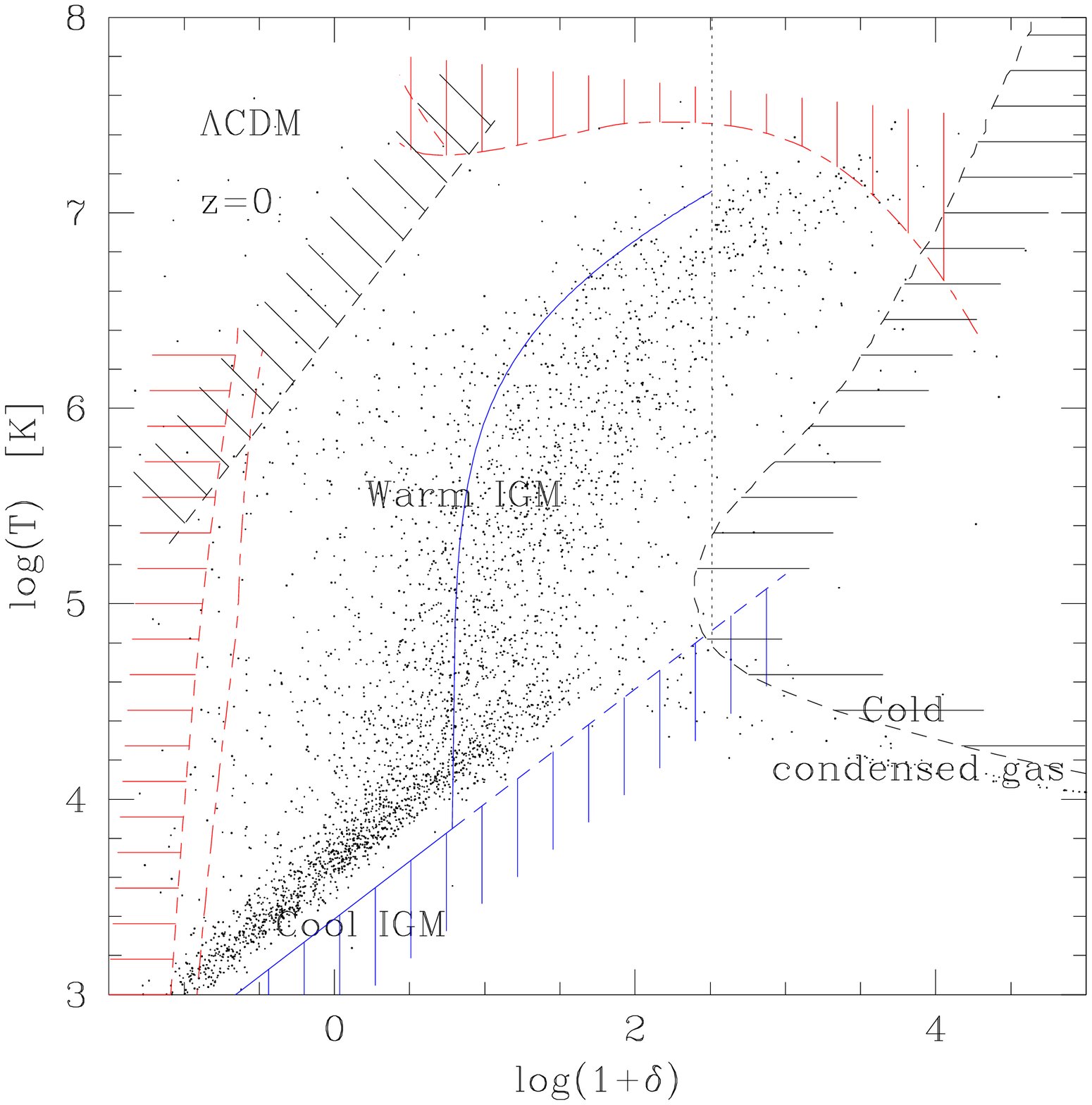}} 
\end{center} 
\caption{The phase-diagram of the IGM from $z=3$ down to $z=0$. The straight solid line $T_{\alpha}$ shows the Equation of State of the ``cool'' IGM (Lyman-$\alpha$ forest). The curved solid line $\Tgh$ shows, as a  mean trend, the ``Equation of State'' of the ``warm'' IGM which is shock-heated through the building of non-linear gravitational structures. The dashed curves draw exclusion regions around the allowed domain for this ``warm'' IGM. In counter-clockwise order, starting from the right side, they correspond to 1) a fast-cooling region where the gas cannot remain over a Hubble time, 2) a high-density and high-temperature domain within the exponential tail of the pdf $\cP(\dR)$ which only contains very rare massive halos, 3) large linear scales where gravitational shock-heating has not appeared yet, 4) a low-density region within the tail of the pdf $\cP(\dR)$ which is associated with very rare voids and 5) a lower-bound for the temperature $T_{\alpha}$ set by radiative heating from the UV background. The vertical dotted line is the density threshold $\Delta_{\rm vir}$ of just-virialized halos. The points show the results of numerical simulations from \cite{Dave1} (Fig.11).} 
\label{diagrhoT}  
\end{figure*}

We display in Fig.\ref{diagrhoT} the Equation of State (\ref{Ta1}) as the solid line which runs through $T \sim 10^4$ K at $\delta = 0$. Beyond the overdensity $\rhoc$ defined in eq.(\ref{dc1}) below, we plot this line as a dashed-line, until it enters the cooling region described in Sect.\ref{Cooling constraint}. Indeed, as we explained in Sect.\ref{High-density fluctuations} at large densities the ``virial temperature'' $\Tloc$ becomes larger than the temperature $\Text$ due to some external energy source (here photo-ionization heating by the UV background radiation). This means that for these regions, which have already reached the non-linear regime as $\delta > \dc \sim 5$ (see eq.(\ref{dc1})), shock-heating due to the gravitational dynamics can no longer be neglected and it actually becomes dominant. Therefore, for high densities with $\rho > \rhoc$ the gas should no longer fall onto the curve (\ref{Ta1}). Nevertheless, since shock-heating can only increase the temperature of the gas the relation (\ref{Ta1}) now provides a lower bound to the temperature $T$. Hence, the region below the curve (\ref{Ta1}) is excluded in the $(\rho,T)$ diagram. This holds until we enter the cooling region discussed in Sect.\ref{Cooling constraint}. On the low-density side, as described in Sect.\ref{Low-density fluctuations} we are constrained by the low-density cutoff $\rhomext(T)$.  
 
Therefore, we predict that we should have two phases for the IGM. A first ``cool'' phase is described by the Equation of State (\ref{Ta1}) with intermediate densities $\rhomext < \rho < \rhoc$. It is  photo-ionized gas heated up to $T \sim 10^4$ K by the background UV flux. This corresponds to the moderate density fluctuations which form the Lyman-$\alpha$ forest. A second ``warm'' phase is made of higher-density regions which have already experienced some shock-heating due to the building of gravitational structures but which have not entered the cooling region yet. These particles should be located in the $(\rho,T)$ plane above the curve (\ref{Ta1}) and within the constraints described in the previous sections.

\subsection{Warm IGM} 
\label{WIGM}

Thus, the previous discussion shows that a ``warm'' component of the IGM should no longer follow the equation of state (\ref{Ta1}). It however remains constrained by the exclusion regions obtained in Sect.\ref{Exclusion constraints}. This leaves a closed allowed region in the $(\rho,T)$ plane. Since the temperature of the gas should be governed by the shock-heating due to the gravitational dynamics we can expect a rather large scatter. Indeed, the history of these particles depends on the properties of their neighbouring dark matter density field. This introduces a stochastic component which gives rise to a wide variety of possible histories so that there is no longer a unique temperature-density relation. This agrees well with Fig.11 in \cite{Dave1} which shows indeed a broad cloud of points in the $(\rho,T)$ plane within the allowed region we obtained in the previous sections. Note also that at $z=0$ some of the gas which has been shock-heated to $T > 10^4$ K actually lies at low density contrasts $\delta <0$ in the numerical simulation. This is consistent with our results shown in Fig.\ref{diagrhoT}. This simply means that on small non-linear scales even regions with $\delta \la 0$ have experienced shell-crossing and gravitational shock-heating (this also implies that they are not just described by a collapse dynamics). 
 
However, the distribution of matter obtained in \cite{Dave1} from numerical simulations does not fill entirely this allowed region (note indeed that this is not implied by our previous considerations) and it is not uniform. Hence it would be convenient to derive a curve in the $(\rho,T)$ plane which would describe the ``mean'' behaviour of this warm phase of the IGM. To this purpose, we first compute the gas temperature associated with a given halo of mass $M$ which forms at redshift $z$. As noticed above, this does not apply to low-density regions with $\delta \la 0$ but this model should provide a useful estimate for the higher density regions which actually contain most of the mass. Gravitational heating is effective in the collapsing phase of the halo which may be set at $\delta > \dc$. One expects $\dc$ to be of the order of a few units. We set it to 
\beq 
1+\dc = 6  \;\;\; \mbox{and} \;\;\;   \rhoc = (1+\dc)\rhob , 
\label{dc1} 
\eeq 
which is close to the value given by the spherical collapse dynamics at turn-around. We consider this not to be an accidental coincidence.  Indeed, the kinetic energy of the underlying overdense large-scale structure is not thermalized during its expansion phase and it only provides an energy source for the heating of the gas within the collapse stage. Even for an expanding cloud which is not perfectly spherical, such a phase of maximum expansion is reached, where the kinetic energy in the collective motion is small. Afterwards, the cloud enters a collapse phase where shocks develop. As is usual for shocks, the latter transform collective large-scale kinetic energy into  small-scale thermal energy. The collective kinetic energy per particle $v^2$ in the collapse phase (i.e. the squared velocity of the inflow) is of order $v^2  \sim  [ \cG M/R -v_0^2]$ where the constant term $v_0^2$ expresses the fact that at turn-around $v^2=0$. Then, this kinetic energy is locally transformed into heat in the shocked regions. The temperature reached there $\Tgh$ is of the order of this kinetic energy since it is set by the inflow velocity $v$ through the matching conditions at the interface between the inner warm shocked region and the outer collapsing region. We thus obtain a relation between the small-scale temperature (which applies to the small shocked regions) and the large-scale kinetic energy per particle. This yields (for a large-scale structure of mass $M$): 
\beq 
\Tgh \sim \Tc \left[ \left(\frac{\rho}{\rhoc}\right)^{1/3} -1 \right] , 
\label{Tgh1} 
\eeq 
with  
\beq 
\Tc = \frac{2}{3} \left(\frac{4\pi}{3}\right)^{1/3} \mu m_p \cG M^{2/3} \rhoc^{1/3} . 
\label{Tc1} 
\eeq 
Of course, the characteristic temperature $\Tc$ depends on the mass $M$ of the collapsing cloud. The subscript ``gh'' refers to the fact that the temperature $\Tgh$ is governed by ``gravitational heating''. This provides an estimate $\Tgh(\rho)$ of the location of the warm spots in the $(\rho, T)$ plane, for fluid elements parameterized by the mass $M$ of the underlying larger-scale structure which is collapsing. Here we assumed that the density within the shocked regions scales as the overall density of the collapsing structure (note that a strong shock only increases the density by a factor 4 at most). 
 
Now, seeking an average location of these warm spots, we estimate the typical mass $M_0(z)$ which collapses at each redshift $z$ by: 
\beq 
\sigma(M_0,z) \equiv 1 . 
\label{Mcol1} 
\eeq 
This also yields the scale $R_0(z)$ defined in eq.(\ref{R01}), with $M_0 = 4\pi/3 \rhob R_0^3$. Among all the $\Tgh(\rho)$ relations (\ref{Tgh1}) which have been seen to depend on the halo mass $M$, we can then find a ``mean'' locus by taking a characteristic temperature $\Tc$, eq.(\ref{Tc1}), calculated with the mass $M_0$. 
 
The mass $M_0$ and the scale $R_0$ characterize the non-linear structures which just turned non-linear. For instance, at $z=0$ we have $R_0 \sim 9$ Mpc. At virialization when the density contrast reaches the threshold $\Dvir \sim 200$ this yields a radius $R_{\rm vir} \sim 1.3$ Mpc as for large clusters. On the other hand, the shocked regions, at temperature $\Tgh$, correspond to much smaller scales. Their size $R$ can be estimated by using approximate hydrostatic equilibrium within the filaments along the transverse direction\footnote{ 
Indeed, the pressure beyond the shock is of order $P \sim \rho \Tgh$ while the velocity drops. The Euler equation then yields $P/\rho \sim \Phi$ where $\Phi$ is the local gravitational potential well associated with the filament. Note that we do not need exact hydrostatic equilibrium within this shocked region: e.g. there may be a non-zero flow along the axis of the filament. We only use the fact that the kinetic energy associated to the transverse collective velocity is small as compared to the local pressure because the shock converts the infall velocity into thermal energy. 
}. 
As a consequence, the eq.(\ref{Tlocp1}) still holds, up to factors of order unity, if $R$ is taken to be the thickness of the filament. This is a natural result since the inflow and the shock actually are at the origin of the filament. Also, the thickness $R$ of the filaments turns out, as a rule, to be much smaller than  $R_0$. For instance, at $z=0$ we find it is of order of a few hundred kpc, see Sect.\ref{Warm IGM two-point correlation} below. Note that the same processes are at work for the dark matter density field, except that shocked regions now correspond to areas where shell-crossing governs the dynamics and builds a large velocity dispersion. Therefore, we can again assume that the gas follows the dark matter density field and the gravitational potential is dominated by the dark matter. Our findings agree with the results of numerical simulations which show that the shocked regions are associated with the filaments which appear in the dark matter density field. Thus, as shown by the simulation map in Fig.3 in \cite{Dave2}, the ``warm'' IGM is an intricate network of filaments which extends over a few Mpc (the scale $R_0$) but the thickness of the filaments (the scale $R$) is much smaller than this global scale.   
In Sect.\ref{Warm IGM two-point correlation} we will see in addition that the scale $R$ of the warm regions shows up as a small-scale cutoff for the ``warm'' IGM two-point correlation function. We obtain a length of a few hundred kpc at $z=0$ which is much smaller than the few Mpc which characterize the underlying global structure which is collapsing. This also agrees with the numerical simulations of \cite{Dave2}. 
 
Therefore, the ``warm'' IGM component is described by the curve (\ref{Tgh1}) in the $(\rho,T)$ plane, together with eq.(\ref{Tlocp1}) which yields the local size $R$ of the clouds (or the thickness of the filaments). Our model relies on three major points: 1) there is a critical overdensity associated to the ``warm'' IGM which reflects the turn-around of patches just going non-linear, 2) shock-heating locally transforms into heat the collective kinetic energy and 3) the scale of the shocked regions is given by local approximate hydrostatic equilibrium at the latter temperature. This yields 1) the locus in parameter space $(\rho,T)$ where the ``warm'' IGM appears, 2) the size of the non-linear network of shocked regions (filament network) which is the scale $R_0$ just turning non-linear and 3) the thickness $R$ of the filaments which appears to be much smaller than the previous scale. 
 
We also see that at the beginning of the cloud collapse the shocked regions are predicted to have low temperatures correlated with a rather small spatial extension. As the overall collapse proceeds their temperature and their size increase and the highest temperatures are reached when the cloud is being virialized with scales of the order of the radius of the halo. This is more or less what could be expected to occur. Thus, we model the mean trend of the ``warm'' IGM by eq.(\ref{Tgh1}) which appears as a curved solid line labeled ``$\Tgh$'' in Fig.\ref{diagrhoT}. It starts from the end-point of the ``Cool'' IGM (at $\delta \sim 5$) and it goes towards larger densities and temperatures until it reaches the cooling region (at $z \ga 3$) or the density threshold $\Delta_{\rm vir}$ (at low $z$). Indeed, beyond this point we consider that we have cooled objects (galaxies) or hot virialized halos (clusters) which are not part of the ``warm'' IGM. 
 
As seen from Fig.\ref{diagrhoT}, at low $z$ gravitational heating rapidly dominates over the UV heating, for densities close to $\rhoc$ and temperatures around $\Tc = \Ta(\rhoc)$. On the other hand, the ``warm'' IGM clouds are not the same as the UV heated objects which form the ``cool'' component. Note that the curve (\ref{Tgh1}) used with the mass (\ref{Mcol1}) only reflects the average trend. Indeed, the stochastic character of the dark matter density field leads to a broad variety of masses which are just collapsing, and hence of $(\rho,T)$ trajectories. Moreover, the local properties of the shocks also provide for some additional scatter. This induces a (rather large) dispersion of the points of the ``warm'' IGM in the $(\rho,T)$ plane, which agrees with numerical simulations (e.g., \cite{Dave2}). However the ``warm'' IGM should remain enclosed within the allowed region defined by the constraints discussed in the previous sections. Indeed, we must point out that the validity of eq.(\ref{Tlocp1}) ensures that the constraints obtained in Sect.\ref{High-density fluctuations} - \ref{Low-density fluctuations} still hold.

\subsection{The different phases of the IGM} 
\label{The different phases of the IGM}

In the previous sections we have shown that one can distinguish two components in the IGM and we have determined their location in the $(\rho,T)$ phase-diagram. Our results are displayed in Fig.\ref{diagrhoT}. 
 
Firstly, there is a ``cool'' IGM phase ($T \sim 10^3 - 10^4$ K) which corresponds to the Lyman-$\alpha$ forest. These are moderate density fluctuations ($\delta < \dc$) of photo-ionized gas. They are described by the Equation of State (\ref{Ta1}) which arises from the heating of the gas by the UV background and the cooling due to the expansion (i.e. pressure work). Thus, this component lies on a well-defined curve in the $(\rho,T)$ plane. This curve is bounded towards low densities by the cutoff of the pdf $\cP(\rhoR)$, which expresses the fact that the dark matter density field arising from Gaussian initial conditions exhibits a finite range of densities which occur with a significant probability. Note that this lower bound indeed agrees with the points obtained from numerical simulations shown in Fig.\ref{diagrhoT}. On the other hand, the high-density bound $\rhoc$ is due to gravitational shock-heating which becomes the dominant energy source for dense regions. 
 
Secondly, there is a ``warm'' IGM phase ($T \sim 10^4 - 10^7$ K) which describes the gas heated by shocks arising from the gravitational energy of just collapsing objects. Because of the stochastic character of this energy source there is a broad scatter for this component around the ``Equation of State'' we have derived. This gas is restricted to a specific allowed region in the $(\rho,T)$ phase-diagram. This expresses cooling and heating constraints as well as the properties of the underlying dark matter density field. We nevertheless obtained a curve, Eq.(\ref{Tgh1}), which follows the mean trend of this ``warm'' phase. Its low density bound is set by the transition near $\rhoc$ with the ``cool'' IGM phase dominated by radiative heating from the UV background. The high-density bound is given by the intersection with the cooling curve (where bremsstrahlung cooling becomes dominant) or the density threshold $\Delta_{\rm vir}$ (beyond this point we have groups or clusters of galaxies). Note that these results are consistent with the calculations of \cite{Nath1} based on the Zel'dovich approximation. 
 
We can note that our results shown in Fig.\ref{diagrhoT} agree reasonably 
 well with the outcome of numerical simulations as displayed in Fig.11 in 
 \cite{Dave1} (also shown by the points in our Fig.\ref{diagrhoT}) and Fig.9 
 in \cite{Spr1}. There is a small offset at low redshift for the 
 normalization of the equation of state (\ref{Ta1}) of the ``cool'' IGM and 
 for the cooling region defined in eq.(\ref{cool1}). Note that the latter 
 could be remedied by adjusting the ratio $\tcoolef/t_H$ which we simply set 
 equal to unity in eq.(\ref{cool1}). Similarly, we could obtain a better fit 
 to the numerical results for the ``warm'' IGM by tuning the r.h.s. in 
 eq.(\ref{Mcol1}) which defines the mass $M_0$. However, our goal is not to 
 get the best fit to a specific numerical simulation (which would be of 
 little value) but to explain the physics of the IGM. Moreover, 
 Fig.\ref{diagrhoT} shows that the simple procedure detailed in the previous 
 sections already provides a good qualitative and quantitative 
 description which should be sufficient for most purposes. Besides, as 
 explained above it should be quite robust. In particular, it could be readily used with any cosmological parameters.

\subsection{Redshift evolution} 
\label{Redshift evolution}

Finally, the four diagrams displayed in Fig.\ref{diagrhoT} show the evolution with redshift of the properties of the IGM, from $z=3$ down to $z=0$. We can see that the pattern does not evolve much qualitatively although the curves exhibit a quantitative shift with $z$. The characteristic temperature of the ``cool'' IGM decreases slightly with time, in agreement with eq.(\ref{Ta1}) since the density declines faster than $1/t_H$. Indeed, the cooling rate due to the expansion of the universe scales as $1/t_H$ while the recombination rate (which yields the density of neutral hydrogen involved in radiative heating) scales as $\rho$. On the other hand, the cooling constraint defined by $\tcoolef = t_H$ in Sect.\ref{Cooling constraint} does not evolve much with time. In particular, its lower branch at $T \sim 10^4$ K is set by the atomic physics of hydrogen ionization. Next, the characteristic temperature of the curve $\rhoploc(T)$ associated with shock-heating due to gravitational clustering grows with time. This expresses the fact that the virial temperature associated with larger scales which turn non-linear later is higher. For instance, the velocity dispersion associated with galaxies is of order $\sim 200$ km/s while for clusters it is $\sim 1000$ km/s. Following this evolution, the mean curve $\Tgh(\rho)$ which describes the ``warm'' IGM enters the cooling region at high $z$ while at low $z$ it first crosses the high-density threshold $\Delta_{\rm vir}$. This actually expresses the fact that at high $z$ the collapsed halos built by gravitational clustering form galaxies since the gas undergoes a very efficient cooling, while at low $z$ typical just-virialized halos are clusters which remain hot over a Hubble time and are still strong X-ray emitters (e.g., \cite{Vclus}).

\subsection{The IGM versus virialized halos} 
\label{virialized halos}

\begin{figure*} 
\begin{center} 
\epsfxsize=8.1 cm \epsfysize=6 cm {\epsfbox{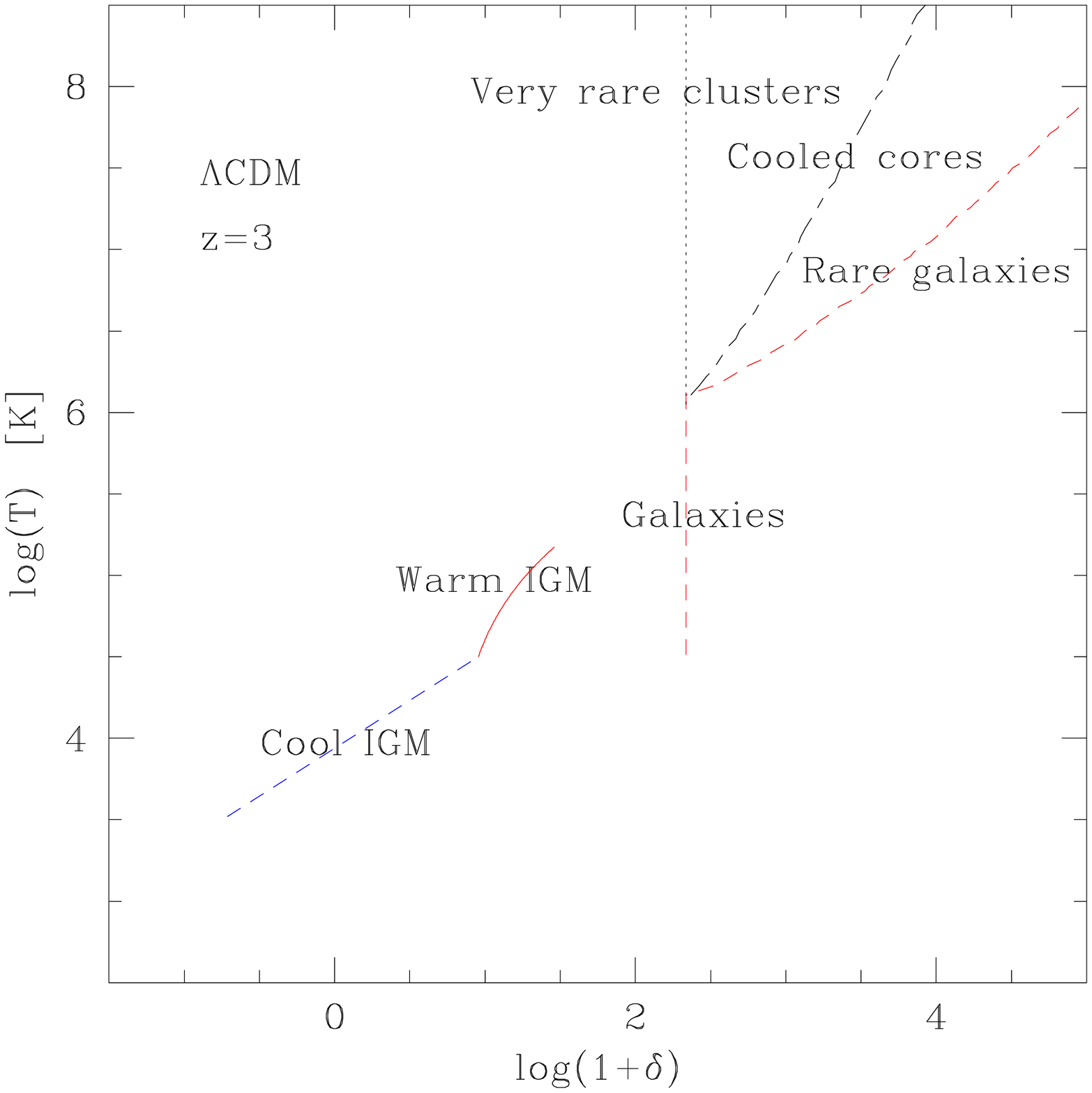}} 
\epsfxsize=8.1 cm \epsfysize=6 cm {\epsfbox{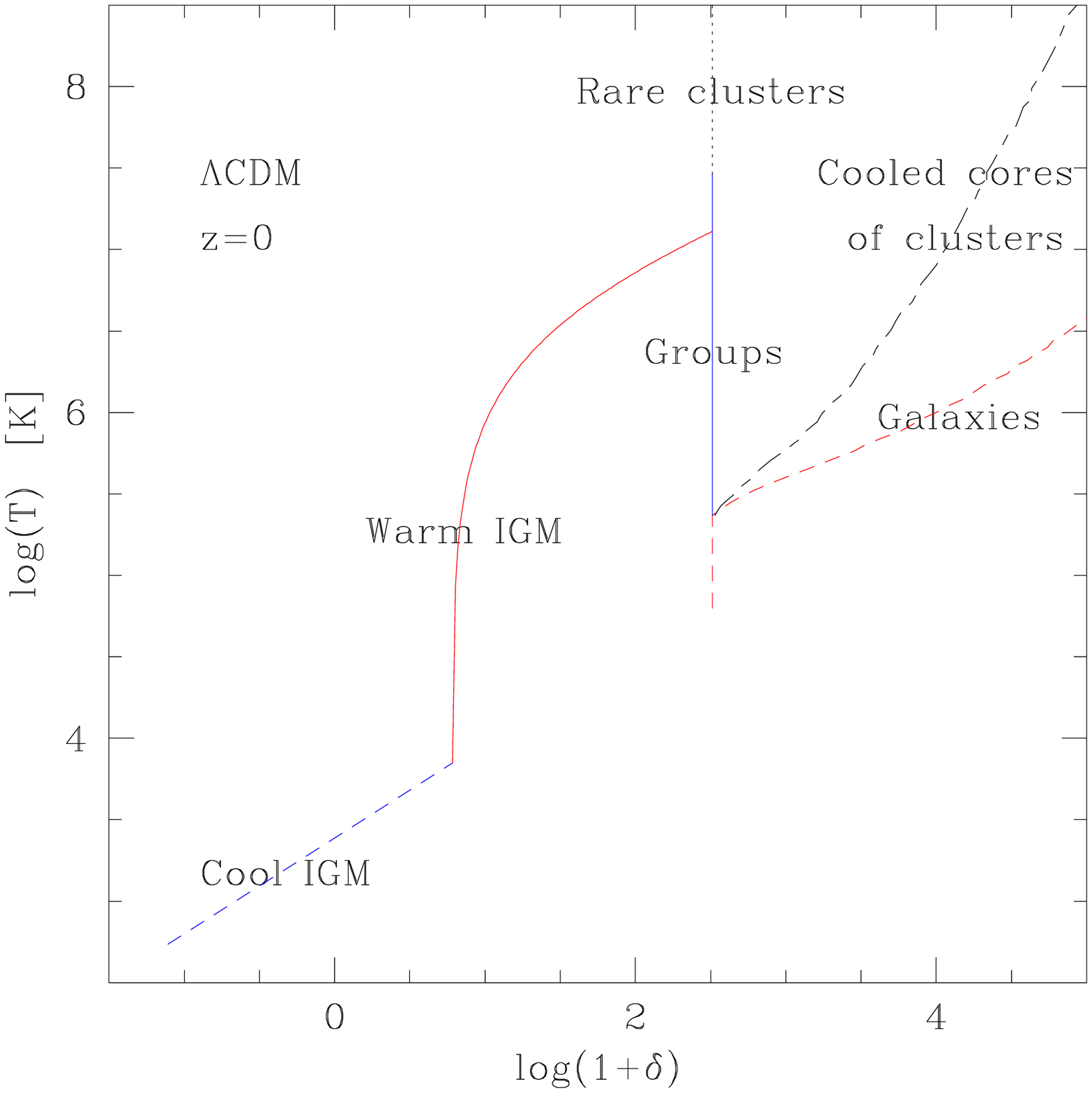}} 
\end{center} 
\caption{The phase-diagram of cosmological baryons at $z=3$ and $z=0$. As in Fig.\ref{diagrhoT} the curves at $\delta < \Dvir(z)$ are the equations of state of the ``cool'' and ``warm'' phases of the IGM. The dashed curve with a vertical part at $\Dvir(z) \sim 200$ and a branch towards higher $\delta$ and $T$ corresponds to galaxies. The two upper parts of the vertical line at $\Dvir(z)$ are groups (solid line at $z=0$) and rare clusters (dotted line). The upper right dot-dashed line corresponds to the cool cores of groups and clusters where cooling has had time to develop.} 
\label{diagtotrhoT}  
\end{figure*}

As explained in the previous sections, we restrict the ``mean equation of state'' of the ``warm'' IGM to density contrasts below the threshold $\Dvir$. Indeed, we consider that larger densities correspond to galaxies or clusters, that is virialized halos which are not part of the diffuse IGM. We display in Fig.\ref{diagtotrhoT} the location in the $(\rho,T)$ phase-diagram of these collapsed objects and of the IGM at redshifts $z=3$ and $z=0$. 
 
Let us first consider the right panel, obtained for $z=0$. The dashed line at $\delta \sim 0$ is the ``cool'' IGM while the solid curve at $\dc < \delta < \Dvir$ is the ``warm'' IGM. The equations of state of these two components were derived in the previous sections and these curves are identical to those shown in Fig.\ref{diagrhoT}. For clarity we do not plot in Fig.\ref{diagtotrhoT} the boundaries discussed in Fig.\ref{diagrhoT} which constrain the large scatter of the ``warm'' IGM phase.  
 
Next, the vertical line at $\delta = \Dvir(z) \sim 200$ shows the overall density contrast associated with just-virialized objects. We divide this line into three parts.  
 
The first is in the low temperatures $T \sim 10^5$ K region. These objects with $\delta = \Dvir$ are located within the cooling region shown in Fig.\ref{diagrhoT}. Hence their cooling time $\tcoolef$ is smaller than the Hubble time $t_H$, see eq.(\ref{cool1}). Therefore, within these halos the gas undergoes a very efficient cooling which leads to the formation of stars. As a consequence, these objects are small galaxies which are just being formed. This part of the vertical line $\delta = \Dvir$ which is enclosed within the cooling region is shown by the lower vertical dashed line in Fig.\ref{diagtotrhoT}. 
 
The remaining high temperature region may still be subdivided. The halos at $\delta = \Dvir$ which have a higher temperature do not cool over a Hubble time since they obey $\tcoolef > t_H$ (except in their center). Hence they correspond to groups or clusters which still contain hot gas which can be observed through its X-ray emission. One may divide this part into two components: groups versus clusters. We identify groups with the low temperature halos located below the cutoff $\rhoploc$ shown in Fig.\ref{diagrhoT}. Therefore, they correspond to the typical just-virialized halos which form at $z=0$. On the other hand, we identify clusters with the high temperature halos located above the cutoff $\rhoploc$. Hence they are rare massive objects which probe the high-density tail of the pdf $\cP(\rhoR)$. In other words, they correspond to the high-mass tail of the mass function. Thus, our distinction between groups and clusters is only based on the abundance of these objects (whether they correspond to rare or typical density fluctuations). However, we shall explain in a future paper (\cite{ValS}) that this subdivision also marks the boundary between objects which, depending on the depth of their gravitational potential, are affected or not by a preheating of the gas through the gravitational processes which yield the IGM. Groups (resp. clusters) are shown in Fig.\ref{diagtotrhoT} by the vertical solid (resp. dotted) line. Thus, these three parts of the vertical line at $\delta = \Dvir$ describe the location of small galaxies, groups and clusters. 
 
For completeness, we must point out that the lower part of the line at $\delta = \Dvir$ with $T \la 2\times 10^5$ K does not give all galaxies. Indeed, it is clear that there are some galaxies with a higher temperature. As discussed in \cite{Vgal} these objects must still satisfy a somewhat stronger cooling constraint: 
\beq 
\tcoolef = t_H(\zform) 
\ , 
\label{cool2} 
\eeq 
where $\zform$ is the redshift of formation of the object. Halos with $\delta=\Dvir(z)$ simply have $\zform=z$ (i.e. they are just being virialized) while higher density contrasts correspond to older objects which collapsed at earlier times. The constraint (\ref{cool2}) ensures that cooling was very efficient when these halos formed so that their gas was able to condense fast enough in order to be left unperturbed by the next generation of forming structures. This allows these objects to retain their individuality, stars and possibly a gaseous disk, even if they become embedded within larger structures (e.g., filaments or clusters) at later times. The condition (\ref{cool2}) at $\delta \geq \Dvir$ is shown by the dashed line in Fig.\ref{diagtotrhoT}, where we used the approximation $t_H(\zform) = t_H \sqrt{(1+\Dvir)/(1+\delta)}$ (i.e. $t_H(\zform)$ scales as the dynamical time of the halo). Thus, we are led to consider that the properties of these galactic halos (mass and size) do not evolve much after they form and reach the virialization density contrast $\delta = \Dvir$. This method yields predictions for the properties of galaxies (luminosity function, masses, metallicity,...) which agree with observations. Therefore, galaxies are defined by the dashed line which shows a first vertical part at $\delta = \Dvir$ and $T \la 2\times 10^5$ K (small galaxies which are being formed) and a second part at higher densities and temperatures (older and more massive galaxies). We can check that we obtain at $T = 10^6$ K a radius $R \sim 50$ kpc which is indeed the typical size of present galaxies. By contrast, a halo with the same virial temperature at $\delta = \Dvir$ would give a size $\sim 300$ kpc (and a mass $\sim 10^{13} M_\odot$) which is uncomfortably large for a galaxy. This actually corresponds to a small group of galaxies. 
 
Finally, we note that the upper part of the cooling curve (\ref{cool1}) shown in Fig.\ref{diagrhoT} at $\delta > \Dvir$ describes the cool cores of groups and clusters. Indeed, it corresponds to high-density regions which have had time to cool today. This yields the core radius of present groups and clusters where cooling has just had time to come into play and possibly induce cooling flows. For instance, at $T = 10^7$ K we obtain a radius $R_{\rm cool} \sim 150$ kpc, which agrees with observations, while the virial radius is $R_{\rm vir} \sim 1$ Mpc. We display this curve as the dot-dashed line in Fig.\ref{diagtotrhoT}. Note that this assumes that baryons follow the dark matter. However, at low $z$ this is not necessarily the case since some ``preheating'' may modify the physics of the gas in low-temperature clusters (i.e. groups), see for instance \cite{VS2} and \cite{Vclus}. This is not important for our purpose here which is mainly to describe the physics of the IGM. A specific study devoted to this problem and the entropy of the gas in the light of the results described in this paper is presented in \cite{ValS}. 
 
Note that for these collapsed objects (galaxies, groups, clusters and cool cores) the density contrast $\delta$ and the temperature $T$ shown in Fig.\ref{diagtotrhoT} only refer to the mean density contrast and virial temperature of the halo over the relevant radius $R$ (which may be different from the virial radius for galaxies and cool cores). At the center of the halo the gas density is larger. Moreover, for galaxies and cool cores the gas temperature can be significantly smaller since cooling is very efficient (and we should take into account feedback from supernovae). Therefore, contrary to the IGM where the location in the $(\rho,T)$ phase-diagram directly gave the properties of the gas, here the  $(\rho,T)$ plane only shows the overall properties of the dark matter halos associated with each object. Finally, we must point out that these classes of objects are not exclusive of each other. Indeed, cooled cores are obviously embedded within groups and clusters while galaxies can be found within filaments, groups and clusters. Therefore, some of the mass associated with galaxies or cool cores is counted within the matter attached to filaments, groups or clusters (this can be handled using the methods of \cite{Vhx}, see \cite{Vgal} for galaxies and \cite{Vclus} for clusters). 
 
Thus, as explained above the $(\rho,T)$ diagram shown in Fig.\ref{diagtotrhoT} describes the physics of cosmological baryons and their distribution between different phases.  
 
The redshift evolution up to $z=3$ can be easily derived from Fig.\ref{diagrhoT} and for illustration we show in the left panel in Fig.\ref{diagtotrhoT} our result at $z=3$. The location of the various curves evolves as explained in Sect.\ref{Redshift evolution} but we can also note some qualitative changes. Firstly, with our definitions we see that there are no more groups. This means that the typical objects which collapse at $z=3$ form galaxies as they exhibit efficient cooling. To be more precise, one can still find clusters along $\delta = \Dvir$, at high temperatures where cooling is not very efficient. However, they are located within the far tail of the mass function and they correspond to extremely rare events. Secondly, we note that the curve which describes the mean ``equation of state'' of the ``warm'' IGM stops below $\Dvir$. Indeed, as seen in Fig.\ref{diagrhoT} its high-density end-point is now given by the intersection with the cooling region. This is merely another consequence of the fact that we typically form galaxies and not groups (this feature coincides with the ``disappearance'' of groups). At an even higher redshift $z \simeq 4$ the ``warm'' IGM component almost disappears as the equation of state of the ``cool'' IGM extends up to the cooling region. However there still remains a ``warm'' phase because the allowed region in the $(\rho,T)$ plane has not vanished\footnote{These objects are not yet virialized and still in their collapse phase: their gas is still expected to undergo shocks which lead to the ``warm'' IGM, but the latter then rapidly cools, so its physics is rather different.}. At these high redshifts most of the mass is within a roughly uniform ``cool'' phase. We still have some rare collapsed halos which correspond to galaxies and some non-linear structures where gravitationally induced shocks heat the gas but these latter regions are severely restricted by the high efficiency of cooling processes.

\section{Distribution of matter} 
\label{Distribution of matter}

\subsection{Mass fraction within different phases} 
\label{Mass within the cool phase}

Thus, in Sect.\ref{The phase-diagram of cosmological baryons} we have 
discussed how the properties of cosmological baryons could be seen through a 
$(\rho,T)$ phase-diagram. In particular, we distinguished a ``cool'' and a 
``warm'' IGM phase. In this section, we derive the fraction of matter 
enclosed within these various components. 
 
Let us first discuss the ``cool'' IGM phase. As described in Sect.\ref{LyEOS} 
this corresponds to the Lyman-$\alpha$ forest, that is moderate density 
fluctuations governed by the ionization and the heating due to the UV 
background. Since its temperature is not zero this gas probes the dark matter 
density field over the scale $R$ defined in eq.(\ref{Rd1}), which describes 
the length-scale over which pressure can homogenize the baryonic matter 
distribution. Note that this length depends on the temperature since $R 
\propto \sqrt{T}$. Then, we wish to express the fraction of matter within the 
``cool'' IGM in terms of the pdf $\cP(\rhoR)$ over the scales $R$ associated 
with these clouds. To do so, we first note that these scales are within the 
non-linear regime as shown by the curve $\Tnl(\rho)$ in Fig.\ref{diagrhoT} 
which marks the transition to the linear regime. Then, as discussed in 
\cite{Vhx}, in the highly non-linear regime the pdf $\cP(\rhoR)$ shows the 
scaling: 
\beq 
\cP(\rhoR) = \frac{1}{\xib^{\;2}} \; h(x) \;\;\; \mbox{with} 
\;\;\; x=\frac{\rho_R}{\xib} , 
\label{hx1} 
\eeq 
which defines the scaling function $h(x)$. This holds for densities 
larger than the low-density cutoffs $\rho_-$ discussed in 
Sect.\ref{Low-density fluctuations}. Note that if the stable-clustering 
ansatz is valid this function $h(x)$ does not depend on redshift, as long as 
one remains in the highly non-linear regime. Then, the fraction of matter 
enclosed within spherical cells of radius $R$ with a density between the 
thresholds $\rho_1$ and $\rho_2$ is: 
\beq 
\Fm = \int_{\rho_1}^{\rho_2} 
\d\rhoR \; \rhoR \; \cP(\rhoR) = \int_{x_1}^{x_2} \frac{\d x}{x} \; x^2 h(x) 
. 
\label{Fm1} 
\eeq 
The advantage of the scaling variable $x$ is that the last term in eq.(\ref{Fm1}) still provides an estimate of the fraction of matter $\Fm$ associated with a curve $(\rho,R)$ which is not necessarily at constant radius, as discussed in \cite{Vhx}. This prediction is compared with numerical simulations in \cite{Lac1} where we investigate the two cases of a constant density and a constant radius. Therefore, in order to estimate the fraction of matter associated with the Lyman-$\alpha$ forest we use the variable $x$ and we write: 
\beq 
\Fm \simeq \int \frac{\d x}{x} \; x^2 h(x;R,z) , 
\label{Fm2} 
\eeq 
with 
\beq 
x^2 h(x;R,z) \equiv \rhoR^2 \cP(\rhoR;R,z) , 
\label{hx2} 
\eeq 
where $\cP(\rhoR;R,z)$ is the value of the pdf $\cP(\rhoR)$ at scale $R$ and redshift $z$ and the integral is taken along the relevant curve $(\rho,R)$. The pdf $\cP(\rhoR)$ is obtained from eq.(\ref{pdf1}) where the parameter $\kappa$ which enters eq.(\ref{zeta1}) is computed at the relevant scale $R$. Thus, the expression (\ref{Fm2}) would be exact within the framework described in App.\ref{The pdf} if the Lyman-$\alpha$ forest were defined by a constant scale $R$. The use of the variable $x$ in eq.(\ref{Fm2}) is only meant to handle the fact that $R$ actually varies with $\rho$. In particular, we must point out that our procedure (\ref{Fm2}) with the definition (\ref{hx2}) does not assume that the stable-clustering ansatz is valid nor that we consider highly non-linear scales. Indeed, the approximation (\ref{Fm2}) merely expresses the mass fraction associated with certain clouds of size $R$ and density $\rhoR$ in terms of the pdf $\cP(\rhoR;R,z)$ measured at this point $\rhoR,R)$.  
 
As described in App.\ref{The pdf} our approximation for the pdf applies both to the linear and non-linear regimes. Moreover, this formalism does not imply that these clouds are virialized objects which have reached an equilibrium state. The ``extended'' function $h(x;R,z)$ defined in eq.(\ref{hx2}) depends on scale and redshift. It is merely another way to write the pdf $\cP(\rhoR;R,z)$ (which goes over to the one relevant for the highly non-linear regime in this limit and is thus used by analogy). An important point of eq.(\ref{Fm2}) is that it yields the mass fraction associated with various objects from the pdf $\cP(\rhoR)$ of the non-linear density field. This is a strong advantage since, as described in \cite{Vhx}, one can show that in the highly non-linear regime, assuming the stable-clustering ansatz is valid, the cloud-in-cloud problem can be handled in a satisfactory way. We can expect this property to extend to the regime we consider here. Note also that eq.(\ref{Fm2}) allows us to count objects which are not necessarily defined by a constant density contrast.

\begin{figure} 
\begin{center} 
\epsfxsize=8.1 cm \epsfysize=6 cm {\epsfbox{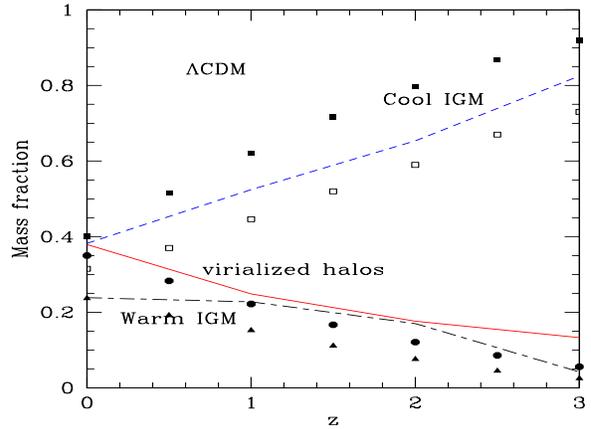}} 
\end{center} 
\caption{The distribution of baryonic matter. The dashed line which increases with redshift is the fraction of matter within the ``cool'' IGM phase, as computed from eq.(\ref{Fm2}) and eq.(\ref{Ta1}). The dot-dashed line gives the mass fraction associated with the ``warm'' IGM component. The solid line is the fraction of matter within collapsed objects. The filled symbols show the results of the numerical simulations from \cite{Dave2} (panel D2 in Fig.1) for the ``cool'' IGM (squares), the ``warm'' IGM (triangles) and condensed gas (circles). The empty squares show the results of the simulation presented in \cite{Dave1} for the ``cool'' IGM phase (Fig.12).} 
\label{frac}  
\end{figure}

Thus, the integration of eq.(\ref{Fm2}) along the Equation of state (\ref{Ta1}) yields the fraction of matter $F_{\rm m,cool}$ within the ``cool'' IGM phase. Next, we could try to use the same method for the ``warm'' IGM, that is to integrate eq.(\ref{Fm2}) along the mean ``equation of state'' (\ref{Tgh1}). However, this procedure is not satisfactory since the filaments and large non-linear structures also contain galaxies and groups which should be subtracted. As a consequence we rather compute the mass fraction associated with virialized halos at $\delta = \Dvir$. That is, we integrate eq.(\ref{Fm2}) along the vertical line at $\Dvir$ shown in Fig.\ref{diagtotrhoT}, starting from the lower point which is given by the lower intersection with the cooling curve. This yields the mass fraction $F_{\rm m,vir}$ associated with virialized halos (galaxies, groups and clusters). Then, we simply define the mass fraction $F_{\rm m,warm}$ within the ``warm'' IGM by: 
\beq 
F_{\rm m,warm} \equiv 1 - F_{\rm m,cool} - F_{\rm m,vir} , 
\label{Fmw} 
\eeq 
so that the sum is unity. This provides the distribution of matter within these three components: the ``cool'' IGM, the ``warm'' IGM and collapsed halos. 
 
We show our results in Fig.\ref{frac}. The Lyman-$\alpha$ forest corresponds to the dashed line which grows at larger redshift. Indeed, at higher $z$ gravitational clustering was less advanced so that a smaller fraction of matter had been shock-heated to high temperatures or embedded within collapsed objects through the building of large scale structures. Note that the fractions of matter within collapsed halos and the ``warm'' IGM are of the same order. Indeed, both components are related to the formation of non-linear gravitational structures. As noticed in Sect.\ref{virialized halos} we obtain $F_{\rm m,warm}=0$ at $z > 4$. This actually means that the mass fraction within the ``warm'' IGM is very small and within the inaccuracy of our computation. 
 
We can check that our results agree with the outcome of the numerical simulations described in  \cite{Dave1} and \cite{Dave2}. The mass fractions we obtain for the ``warm'' IGM and virialized objects are also in reasonable agreement with simulations although it is difficult to make a detailed comparison. Indeed, our separation between different components is not exactly the same as in the simulations\footnote{We define, in what we consider a somewhat more logical way, the various components according to their equation of state in the phase-diagram, which is easily recognizable for the ``cool'' or ``warm'' IGM, or for the virialized halos. This definition, as expected, changes accordingly with redshift. In the simulations, on the other hand, a simple temperature cut-off is used. For the components containing a small fraction of the baryons, this may introduce some difference.} and the predictions of the various simulations exhibit a significant scatter as shown by the comparison between the filled and empty squares in Fig.\ref{frac} (see also the various panels in Fig.1 in \cite{Dave2}).

\subsection{Temperature distribution of the warm phase} 
\label{Temperature distribution of the warm phase}

\begin{figure} 
\begin{center} 
\epsfxsize=8.1 cm \epsfysize=6 cm {\epsfbox{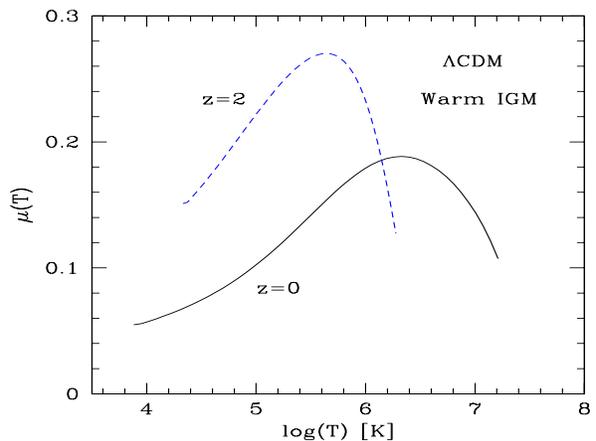}} 
\end{center} 
\caption{The temperature distribution $\mu(T)$ of the ``warm'' IGM component at redshifts $z=2$ and $z=0$ (normalized to unity). The mean temperature grows with time as larger scale structures form.} 
\label{fracT}  
\end{figure}

We have shown in Fig.\ref{frac} the fraction of matter within the ``warm'' IGM. It is also of interest to consider more precisely the temperature distribution within this ``warm'' IGM. As we explained in Sect.\ref{Mass within the cool phase}, the mass function which describes the ``warm'' IGM cannot be obtained by directly integrating eq.(\ref{Fm2}) along the mean ``equation of state'' (\ref{Tgh1}) because the regions which are about to collapse also contain smaller virialized halos (galaxies, groups). Thus, we first need to subtract these substructures which yields eq.(\ref{Fmw}) for the mass fraction. This subtraction is more difficult when we consider the detailed temperature distribution within the ``warm'' IGM but we shall assume that it can be described by a simple overall renormalization. This would be exact if the fraction of matter enclosed within such virialized objects would be the same for all ``warm'' IGM regions (i.e. filaments). In fact, we can expect some correlation between these substructures and the larger-scale filaments but this procedure can be seen as a zeroth-order approximation. As a consequence, following eq.(\ref{Fm1}) we write the fraction of matter $\mu(T) \d T/T$ within the range $[T,T+\d T]$ as: 
\beq 
\mu(T) \frac{\d T}{T} \propto x^2 h(x) \frac{\d x}{x} , 
\label{muT1} 
\eeq 
which we compute along the curve (\ref{Tgh1}) which describes the mean ``equation of state'' of the ``warm'' IGM. In eq.(\ref{muT1}) and in the following we do not write explicitly the time and scale dependence of $h(x;R,z)$. The normalization constant in eq.(\ref{muT1}) is set by the constraint $\int \mu(T) \d T/T = F_{\rm m,warm}$, which is the total mass fraction within the ``warm'' IGM derived in Sect.\ref{Mass within the cool phase}. 
 
We display our result normalized to unity in Fig.\ref{fracT}. Of course, we recover the fact that the mean temperature increases with time as larger non-linear structures build up. This could also be seen from Fig.\ref{diagrhoT}. In particular, the mean trend curve of the ``warm'' IGM in Fig.\ref{diagrhoT} at $z=0$ is very steep because the characteristic temperature of the non-linear structures which collapse at that time is of order $\Tcol \sim 2 \times 10^7$ K which is much larger than the temperature reached by radiative heating from the UV background $T_{\alpha} \sim 5 \times 10^3$ K. Therefore, gravitational shock-heating easily heats the gas up to $T \ga 10^6$ K. However, note that there is still a significant fraction of the matter at lower temperatures $T < 10^6$ K, in agreement with \cite{Dave2}. Although the procedure (\ref{muT1}) is only a simple approximation, the comparison of Fig.\ref{fracT} with Fig.5 in \cite{Dave2} shows that it captures the main trend. Of course, as explained in Sect.\ref{WIGM} the ``warm'' IGM shows a broad scatter in the $(\rho,T)$ plane so that the actual boundaries of the temperature distribution are not as sharp as in Fig.\ref{fracT}. On the other hand, note that in our description we also have some matter at $T>10^7$ K but we associate this gas with massive X-ray clusters. In any case, our result displayed in Fig.\ref{fracT} appears to provide a reasonable description of the ``warm'' IGM properties.

\section{Warm IGM two-point correlation} 
\label{Warm IGM two-point correlation} 
 
Another check on our description of the ``warm'' IGM component is provided by its two-point correlation function $\xiw(r)$ which measures its spatial clustering. Hence in this section we evaluate $\xiw(r)$, which is defined by: 
\beq 
1+\xiw(\br) \equiv \frac{\lag \rhow(\br_1) \rhow(\br_1+\br) \rag}{\lag \rhow \rag^2} , 
\label{xiw1} 
\eeq 
where $\rhow(\br)$ is the density of the ``warm'' IGM component at point $\br$. In eq.(\ref{xiw1}) the averages $\lag .. \rag$ are over the realizations of the density fields or over space (assuming ergodicity). First, following the procedure described in Sect.\ref{Temperature distribution of the warm phase} we can write the mean ``warm'' IGM density as: 
\beq 
\lag \rhow \rag = \cN \int \frac{\d x}{x} x^2 h(x) \frac{\rhob}{M} V \rho = \rhob \Fmw  
\label{rhow1} 
\eeq 
where $M$, $R$ and $V$ are the mass, the radius and the volume associated with the point $(\rho,T)$ in the phase-diagram. The normalization constant $\cN$ is defined by: 
\beq 
\cN \int \frac{\d x}{x} x^2 h(x) \equiv \Fmw . 
\label{N1} 
\eeq 
Note that we always have $\lag \rhow \rag < \rhob$ since the ``warm'' IGM component does not contain all the baryonic matter. Next, the second-order moment is given by: 
\beq 
\lag \rhow(\br_1) \rhow(\br_1+\br) \rag = \cN \rhob \int \frac{\d x_1}{x_1} x_1^2 h(x_1) \rhow(\br,x_1) 
\label{xiw2} 
\eeq 
where $\rhow(\br,x_1)$ is the mean ``warm'' IGM density at distance $\br$ from a point embedded within the ``warm'' IGM component with the parameter $x_1$. Then, as in \cite{Silk2}, we write $\rhow(\br,x_1)$ as the sum over two contributions. First, there is a probability $F(r/R_1)$ that the second point at $\br_1+\br$ belongs to the same region of thickness $R_1$ of the ``warm'' IGM. Assuming spherical regions we have: 
\beq 
r < R_1 : F = 1 - \frac{3 r}{2 R_1} + \frac{r^3}{2 R_1^3} , \; r> R_1 : F = 0 . 
\label{F1} 
\eeq 
This gives the first contribution $\rhow^{(1)}(\br,x_1)$ as $\rhow^{(1)}(\br,x_1) = \rho_1 F(r/R_1)$. Second, we write the contribution $\rhow^{(2)}(\br,x_1)$ to $\rhow(\br,x_1)$ of other regions of the ``warm'' IGM component with parameter $x_2$ as: 
\beqa 
\rhow^{(2)}(\br,x_1) & = & \left[ 1 - F(r/R_1) \right] \cN \rhob \int \frac{\d x_2}{x_2} x_2^2 h(x_2) \nonumber \\ & & \times \left[ 1 + b(x_1) b(x_2) \xi(r) \right] . 
\label{rhow21} 
\eeqa 
Here $\xi(r)$ is the two-point correlation of the underlying dark matter density field while $b(x_1) b(x_2)$ is the bias associated with the regions $x_1$ and $x_2$. This bias is computed from the generating function introduced in eq.(\ref{pdf1}) as described in \cite{Ber3} (see also \cite{Val4} for a comparison with observations). Thus, we write the second-order moment as: 
\beqa 
\lefteqn{ \lag \rhow(\br_1) \rhow(\br_1+\br) \rag = \cN \rhob^{\;2} \int \frac{\d x_1}{x_1} x_1^2 h(x_1) \biggl \lbrace \rho_{R1} F(r/R_1) } \nonumber \\ & & + \left( 1 - F(r/R_1) \right) \cN \int \frac{\d x_2}{x_2} x_2^2 h(x_2) \left[ 1 + b(x_1) b(x_2) \xi(r) \right] \biggl \rbrace \nonumber \\ 
\label{xiw3} 
\eeqa 
This yields the ``warm'' IGM two-point correlation $\xiw(r)$ through eq.(\ref{xiw1}).

\begin{figure} 
\begin{center} 
\epsfxsize=8.1 cm \epsfysize=6 cm {\epsfbox{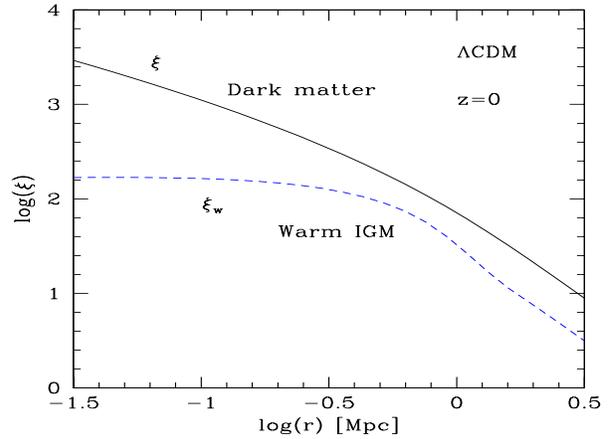}} 
\end{center} 
\caption{The ``warm'' IGM two-point correlation $\xiw(r)$ at redshift $z=0$ (dashed line). The solid line shows the two-point correlation $\xi(r)$ of the dark matter density field.} 
\label{XiW}  
\end{figure}

We show our result at $z=0$ in Fig.\ref{XiW}. On large scales the ``warm'' IGM correlation closely follows the dark matter correlation because i) we assumed that baryonic matter follows the dark matter density field smoothed over the scale $R$ which describes the processes at work ($R$ is set by the location on the $(\rho,T)$ plane) and ii) the ``warm'' IGM component is mainly made of typical density fluctuations. This is seen by the fact that, by definition, most of its matter is within the allowed region in the $(\rho,T)$ phase-diagram, below the high-density cutoff $\rhoploc(T)$. This implies that the biases which enter the expression (\ref{xiw3}) are close to unity since we do not select rare high densities or voids by looking at the ``warm'' IGM phase. In fact, as seen from Fig.\ref{XiW} we obtain a mean bias which is slightly smaller than unity. However, this does not take into account the broad scatter in the $(\rho,T)$ plane of the ``warm'' IGM component. 
 
On the other hand, at small scales $r \la 300$ kpc the two-point correlation function $\xiw(r)$ flattens and it reaches a finite limit at $r=0$. This expresses the fact that the non-zero temperature of the gas reached by radiative heating and shock-heating ($T > 10^4$ K) gives rise to pressure effects which homogenize the baryonic distribution over some scale $R$. A lower limit to this length $R$ is set by the scale $R_c$ associated with the critical point $(\rhoc,\Tc)$ in the phase-diagram. Indeed, as shown by the curved solid line in Fig.\ref{diagrhoT}, at redshift $z=0$ the length scales associated with the ``warm'' IGM run from $R_c$ up to scales of order $R_0$ which are turning non-linear. This sets the knee of the two-point correlation $\xiw(r)$ at $r \sim 300$ kpc at $z=0$. As noticed in Sect.\ref{WIGM}, the result displayed in Fig.\ref{XiW} shows that the characteristic thickness of the filaments which build the ``warm'' IGM is of order $300$ kpc at $z=0$. We find this, at a given redshift, to be indeed much smaller than the scale which is just turning non-linear, in agreement with numerical simulations.  
 
The value at $r=0$ yields the clumping factor $\Cw$ which we define by: 
\beq 
\Cw \equiv \frac{\lag \rhow^2\rag}{\lag \rhow\rag^2} = 1+\xiw(0) . 
\label{Cw1} 
\eeq 
We obtain $\Cw \simeq 160$ at $z=0$. Finally, we can check that our results agree with the outcome of the numerical simulations described in \cite{Dave2} (see their Fig.7). In particular, we recover the knee at $r \sim 300$ kpc and a reasonable value for the clumping factor $\Cw$. Note however that the dispersion in the $(\rho,T)$ phase-diagram may actually lead to a slightly higher value.

\section{Soft X-ray background} 
\label{Soft X-ray background}

Finally, we check in this section whether the soft X-ray background emitted by the ``warm'' IGM in our model is consistent with observations. Experiments show that the extragalactic soft X-ray background flux in the $0.1-0.4$ keV band is of order $20-35$ keV cm$^{-2}$ s$^{-1}$ sr$^{-1}$ keV$^{-1}$ (e.g., \cite{War1}). However, most of this flux is due to AGN so that the contribution from diffuse gas should obey the constraint $F_{\nu} \la 4$ keV cm$^{-2}$ s$^{-1}$ sr$^{-1}$ keV$^{-1}$ in this frequency range (e.g., see the discussion in \cite{Wu1}). Using the Press-Schechter formalism (\cite{PS1}), assuming that the X-ray emitting gas is embedded within just-virialized halos with a density contrast $\Delta_c(z) \ga 177$, some previous studies obtained a flux which is higher than this upper bound (e.g., \cite{Pen1}, \cite{Wu1}). This led these authors to infer that a non-gravitational heating source is needed to unbind the gas from groups and small clusters in order to reduce their contribution to the soft X-ray background. However, the diffuse ``warm'' IGM component is made of filamentary structures with density contrasts lower than $\Delta_c(z)$. This gives a smaller X-ray background since the X-ray emission is proportional to the squared density of the gas. Moreover, the fraction of matter within this ``warm'' diffuse phase is smaller than $30\%$, see Fig.\ref{frac}. As argued in \cite{Dave2} this reduced X-ray background could then be consistent with observational constraints. This conclusion also agrees with the outcome of numerical simulations (\cite{Croft1}, \cite{Phil1}). 
 
Therefore, we compute here the soft X-ray background due to the diffuse ``warm'' IGM phase, in the $0.1-0.4$ keV band. Note that the X-ray background due to resolved sources (AGN, clusters, cooling galaxies) was already studied in \cite{Vclus} from an analytic model similar to the one used here to describe the underlying dark matter density field. The X-ray flux $F_{\nu_1-\nu_2}$ within this frequency band can be written: 
\beqa 
\lefteqn{ F_{\nu_1-\nu_2} = 7.13 \times 10^{-34} \; \mbox{keV cm$^{-2}$ s$^{-1}$ sr$^{-1}$} } \nonumber \\ & & \times \int\d z \; c \frac{\d t}{\d z} (1+z)^2 \cN(z) \int \frac{\d x}{x} x^2 h(x) (1+\delta) T^{1/2} \nonumber \\ & & \times \left( e^{-h \nu_1 (1+z)/(kT)} -   e^{-h \nu_2 (1+z)/(kT)} \right) , 
\label{FX1} 
\eeqa 
where the temperature $T$ is in Kelvin units. Here we used again eq.(\ref{muT1}) with the normalization $\cN(z)$ defined in eq.(\ref{N1}). The factor $(1+\delta)$ in eq.(\ref{FX1}) comes from the fact that the X-ray emission is proportional to the squared density $n_e^2$. The integral over $x$ in eq.(\ref{FX1}) follows the curve $\Tgh(\rho)$, from eq.(\ref{Tgh1}), in the $(\rho,T)$ phase-diagram. Then, our calculation gives an X-ray background flux $F_{\nu_1-\nu_2} = 0.07$ keV cm$^{-2}$ s$^{-1}$ sr$^{-1}$ in the $0.1-0.4$ keV band. This yields a mean differential flux $F_{\nu} = 0.25$ keV cm$^{-2}$ s$^{-1}$ sr$^{-1}$ keV$^{-1}$ in this band. Thus, we can check that our result is consistent with observations. Note that the scatter of the ``warm'' IGM in the $(\rho,T)$ plane should lead to a slightly larger flux. Therefore, this component might be observable through its X-ray emission in the near future.

\section{Conclusion}

In this article we have shown that the $(\rho,T)$ phase-diagram of cosmological baryons can be understood in very simple terms. It is strongly model-independent since the relevant scales are set by atomic physics (hydrogen ionization) and the basic cosmological setting (the rate of expansion and the scale which marks the transition to the non-linear regime). Building a bridge between the modeling of galaxies, groups and clusters, as well as Lyman-$\alpha$ absorbers, we have constructed in this paper a phase-diagram which includes also the IGM, and thus provides a {\it phase-diagram of all baryons}. In this diagram, all these objects form well-defined populations.

Firstly, we have distinguished a ``cool'' IGM phase corresponding  
to the Lyman-$\alpha$ forest. It follows a well-defined equation of state on 
the $(\rho,T)$ plane. We have shown that the properties of this gas do not 
depend much on its previous history because of the form of the recombination 
coefficient. This explains why the scatter of this component obtained in 
numerical simulations is quite small. The fraction of baryons in this components, as well as its evolution with redshift is seen to be consistent with earlier modelling of the frequency of occurrence of the Lyman-$\alpha$ lines.   We however obtain (consistently with the numerical simulations) 38\% 
of the baryons in the cool IGM at z=0. 
This may be a little high with  
a mass fraction closer to ~30\%  
being  consistent with observations. 
This must be considered as being within the error bars of our analytical calculation (and the error bars of the simulations!).

Secondly, we find that a ``warm'' IGM phase is formed by the gas which has 
 been shock-heated to larger temperatures $T \sim 10^4 - 10^7$ K as non-linear 
 gravitational structures appear. The dependence on the stochastic 
 gravitational potential entails a broad scatter in the $(\rho,T)$ 
 plane. However, we have explained that robust constraints only allow a 
 closed region in the $(\rho,T)$ phase-diagram for this component. We have 
 also defined a simple curve which represents its mean behaviour and plays 
 the role of the ``warm'' IGM Equation of State. The latter is found to be 
 consistent with the outcome of numerical simulations (\cite{Dave2}), but 
 differs from the simple power-law fit of a mean curve done there. This is 
 because its locus is based on physical considerations, which we have 
 thoroughly justified. The ``warm'' IGM is seen to be due to large structures 
 on their way of collapsing under the action of gravity.  The quite simple 
 picture behind this model is that the collapse   of an object turning 
 non-linear (that is with average overdensity above $5$) induces shocks which 
 heat patches of much smaller size, to a temperature of the order of the 
 locally available kinetic energy of the collapse, creating local conditions within hot spots  
 close to hydrostatic equilibrium. We are able to estimate this size, as well 
 as the temperature of the latter, and to show they are in agreement with the 
 simulations. Definitely, these objects are not just Lyman-{$\alpha$} 
 absorbers at larger densities. The hot gas which lies 
 within clusters and galactic halos, on the other hand, corresponds to the high-density 
 ``continuation'' (e.g., $\delta \ga 200$) of the ``warm'' IGM curve, in a 
 region of phase-space where virialization insures that all the available gas 
 has been gravitationally heated. 
 
Next, we find two quite different regimes for the ``warm'' IGM. At high 
redshift, some of the ``warm'' component enters the cooling region of the 
$(\rho,T)$ plane: this gives rise to galactic disks and stars. 
Some of the baryonic
matter, also, should lie on the low temperature branch of this cooling region, 
in agreement with the results of numerical simulations (\cite{Dave1}). At low 
redshift, the ``warm'' IGM no longer goes into the cooled region, 
and it provides 
the origin of the hot gas in clusters. {\it The evolution with redshift of 
the baryon phase-diagram provides a natural confirmation that at high $z$ 
collapsed halos form galaxies while at low redshift they build groups or 
clusters}. The warm phase occupies 24\% 
of the baryon fraction at $z=0$ in 
our analytic model. 
 
Then, we have checked that our results for the fraction of matter enclosed 
within the various phases and the two-point correlation function of the 
``warm'' IGM component agree with numerical simulations (e.g., \cite{Dave1} 
and \cite{Dave2}). This confirms the validity of our analysis. Note however 
that the latter relies on simple physical considerations and it is 
independent of the findings of the former. Besides, the soft X-ray background 
due to the ``warm'' IGM is consistent with the upper bound set by 
observations. Our prediction actually is not far from the observational 
limit: this offers the prospect of measuring this X-ray emission in future 
observations. 

We predict that approximately 60\% 
of the baryons are accounted for at z=0 in the cool/warm phases of the intergalactic medium. The remaining baryons (i.e. 40\%) 
are in collapsed structures. With a total of $\Omega_b \approx 0.045$, this amonts to $\Omega_{\rm bcoll} \approx 0.018$.
The observed luminous component of the baryons (stars, remnants and gas) is estimated to be about $\Omega_{\rm gal} \approx 0.004 $, as noted in the introduction, and the hot gas in clusters represents $\Omega_{\rm ghot} \approx 0.003$.  Therefore, there remains an ``unidentified'' or dark baryon component,
$\Omega_{\rm bdm} \approx 0.011$. This is at least twice the stellar-like component but only about 1/4 of the ``identified''  baryons (IGM, stellar components and very hot gas). In our model this dark baryonic matter corresponds to (cool ?) gas within galactic halos and groups, which has not been observed yet.

\begin{acknowledgements} 
We thank R. Dave for providing us with some of the data published in \cite{Dave1}. 
\end{acknowledgements}

\appendix 
 
\section{Dark matter density field} 
\label{Dark matter density field} 
 
In this appendix we describe the analytic procedure we use to model the dark matter density field. Our method provides a simple means to estimate the pdf of the density contrast from the quasi-linear up to the highly non-linear regime, parameterized by only one variable (the skewness). This should be sufficient for our purpose which is merely to obtain the main properties of the IGM. Note that by construction the variance and the skewness of the pdf we obtain below agree with the results of numerical simulations, in both the quasi-linear and the highly non-linear regimes.

\subsection{The pdf $\cP(\rhoR)$ of the overdensity at scale $R$} 
\label{The pdf}

First, we note $\xib$ the moment of order two of the density contrast $\dR$ over a spherical cell of radius $R$ and volume $V$. It can be expressed in terms of the two-point correlation function $\xi(\bx_1,\bx_2)=\xi(|\bx_1-\bx_2|)$ as: 
\beq 
\xib \equiv \lag \dR^2 \rag = \int_V \frac{\d\bx_1 \d\bx_2}{V^2} \; \xi(\bx_1,\bx_2) . 
\label{xi1} 
\eeq 
The two-point correlation $\xi(x)$ is the Fourier transform of the non-linear power-spectrum $P(k)$ which we compute from the linear power-spectrum $P_L(k)$ using the analytic formulae obtained by \cite{Peac1} from fits to N-body simulations. Of course, in the quasi-linear limit $\xib \rightarrow 0$ we have $\xib=\sigma^2$. Then, it is convenient to introduce the generating function $\varphi(y)$ defined by the inverse Laplace transform: 
\beq 
\cP(\rhoR) \equiv \inta \frac{\d y}{2\pi i \xib} \; e^{[\rhoR y - \varphi(y)]/\xib} , 
\label{pdf1} 
\eeq 
where $\cP(\rhoR)$ is the pdf of the overdensity over a cell of radius $R$: 
\beq 
\rho_R \equiv 1+\dR . 
\label{rhoR1} 
\eeq 
The advantage of the generating function $\varphi(y)$ is that it reaches a finite limit in the quasi-linear regime $\xib \ll 1$. Moreover, in this regime one can derive $\varphi(y)$ in a rigorous manner from the equations of motion (Bernardeau (1992,1994), \cite{ValII}). On the other hand, in the highly non-linear regime $\xib \gg 1$ the generating function $\varphi(y)$ should also reach a finite (astonishingly close to the former, but seemingly not identical) limit if the stable-clustering ansatz is valid (\cite{Bal1}). This is seen to agree reasonably well with the results of numerical simulations (e.g., \cite{Lac1}). Therefore, hereafter we use the following simple parameterization for the generating function $\varphi(y)$: 
\beq 
\left\{ \begin{array}{l} 
{\displaystyle \tau = - y \; \zeta'(\tau) } \\ \\ 
{\displaystyle \varphi(y) = y \; \zeta(\tau) + \frac{\tau^2}{2} } 
\end{array} \right. 
\label{phitau1} 
\eeq 
with: 
\beq 
\zeta(\tau) \equiv \left( 1 + \frac{\tau}{\kappa} \right)^{-\kappa} . 
\label{zeta1} 
\eeq 
This yields a family of functions $\varphi(y)$ which  depends only  on the parameter $\kappa$.

Such a relationship arises naturally in the exact derivation of $\varphi(y)$ in the quasi-linear limit (Bernardeau (1992,1994), \cite{ValII}). Besides, it also appears in the non-linear regime within the framework of the ``minimal tree-model'' (\cite{Ber3}, \cite{Sch1}). It can be seen that the phenomenological form (\ref{zeta1}) for $\zeta(\tau)$ also yields reasonable results.

As in \cite{Ber4}, we choose the parameter $\kappa$ which enters eq.(\ref{zeta1}) so as to recover the skewness $S_3 \equiv \lag \dR^3 \rag_c/\xib^2$ predicted by HEPT (\cite{Scoc1}) in the non-linear regime. More precisely, we use: 
\beq 
\xib < 0.1 : \; S_3 = \frac{34}{7} - (n+3) , \;\; \xib > 10 : \; S_3 = \frac{3(4-2^n)}{1+2^{n+1}} 
\label{S31} 
\eeq 
and we interpolate between both values in the range $0.1<\xib<10$. Here we note $n$ the slope of the linear-power spectrum at the scale of interest. Note that this procedure was seen in \cite{Ber4} to match the results of numerical simulations with respect to weak gravitational lensing effects. Therefore, it should be quite sufficient for our present purpose which is mainly to investigate the main trends of the IGM properties.

\subsection{High and low density cutoffs $\rhopm(R)$} 
\label{Density cutoffs}

Now, we can estimate the high and low density cutoffs $\rhopm(R)$ of the pdf $\cP(\rhoR)$ as follows. To each overdensity $\rhoR$ we associate the Laplace variable $y_c(\rhoR)$ (and $\tau_c$ through eq.(\ref{phitau1})) defined as the saddle-point of the exponent in eq.(\ref{pdf1}). This can be written: 
\beq 
\rhoR \equiv \varphi'(y_c) = \zeta(\tau_c) , 
\label{yc1} 
\eeq 
where we used eq.(\ref{phitau1}) to derive the last equality. Note that $\dR$ and the pair $(y_c, \tau_c)$ are of opposite signs. The reason for the introduction of $y_c$ is that for extreme overdensities $\rhoR \rightarrow \infty$ or $\rhoR \rightarrow 0$, that is in the tails of the pdf $\cP(\rho_R)$, the integral (\ref{pdf1}) is dominated by the contribution from $y \simeq y_c$ (e.g., \cite{Bal1}). Then, it is clear that the density cutoffs of the pdf are given by the points $\rhopm$ where the argument of the exponential in eq.(\ref{pdf1}) at the saddle-point is of order unity and negative (note that it is zero at the mean overdensity $\lag \rhoR \rag=1$). Indeed, for larger $|\dR|$ the argument becomes more negative than $-1$ which leads to the exponential falloffs of $\cP(\rhoR)$. Therefore, we define the high and low density cutoffs $\rhopm$ by the condition: 
\beq 
\frac{\rhopm \ypm - \varphi(\ypm)}{\xib} = -1 , 
\label{rhpm1} 
\eeq 
where $\ypm=y_c(\rhopm)$ as given by eq.(\ref{yc1}). This relation can be most conveniently written in terms of the variable $\tau$ which yields: 
\beq 
\tau_{\pm} = \mp \sqrt{2 \xib} 
\label{taupm1} 
\eeq 
where we used eq.(\ref{zeta1}). In the highly non-linear regime eq.(\ref{taupm1}) yields $\tau_+ \ll -1$. However, as discussed in \cite{Ber3} the implicit system (\ref{phitau1}) is singular at a finite value $\tau_s = - \kappa /(2+\kappa)$ which leads to an exponential tail for the pdf $\cP(\rhoR) \sim e^{-\rhoR/(x_s \xib)}$. Hence for $-\sqrt{2 \xib}<\tau_s$ we take $\tau_+=\tau_s$ and $\rhop=x_s \xib$. Thus, we obtain for the high-density cutoff $\rhop(R)$: 
\beq 
\xib \ll 1: \; \rhop = 1 + \sqrt{2 \xib} , \;\;\;\; \xib \gg 1: \; \rhop = x_s \xib 
\label{rhopR1} 
\eeq 
and for the low-density cutoff $\rhom(R)$: 
\beq 
\xib \ll 1: \; \rhom = 1 - \sqrt{2 \xib} , \;\;\;\; \xib \gg 1: \; \rhom \sim \xib^{\;-\kappa/2} . 
\label{rhomR1} 
\eeq 
Note that in the quasi-linear regime we recover $|\delta_{\pm}| \sim \sqrt{\xib}=\sigma$ as required for Gaussian initial conditions. Then, the density cutoffs $\rhopm$ are close to the mean: $\rhopm \sim \lag \rhoR \rag=1$ for $\xib \ll 1$. On the other hand, in the highly non-linear regime the cutoffs are displaced to very low densities ($\rhom \ll 1$: rare voids) and high densities ($\rhop \gg 1$: collapsed halos).

\end{document}